\documentclass{aa}
\usepackage{txfonts}
\usepackage{graphicx}

\usepackage{natbib}
\bibpunct{(}{)}{;}{a}{}{,}

\def\cd  {{$\mbox{c~d}^{-1}$}}
\def\kms {{$\mathrm{km}\,\mathrm{s}^{-1}$}}
\def\vsini {{$\mathrm{V}\!\sin\!i$}}
\def\omc{$\Omega/\Omega_{\rm{c}}$}

\def\teff{T$_{\rm eff}$}

\begin{document}

\title{A photometric study of Be stars located in the seismology fields of COROT\thanks{Section~5, including Figs.~\ref{fig:42259:1}-\ref{fig:183656:1} and Tables~\ref{table:summary:anti}-\ref{table:summary:cen}, are only available in electronic form at {\tt http://www.aanda.org}}
}

\author{J. Guti\'errez-Soto\inst{1,2}, J. Fabregat\inst{1,2}, J. Suso\inst{3}, M. Lanzara\inst{3}, 
R. Garrido\inst{4}, A-M. Hubert\inst{2} and M. Floquet\inst{2}  }

\offprints{juan.gutierrez-soto@uv.es}
\institute{
Observatorio Astron\'omico. Universidad de Valencia.
    Edicifio Institutos de Investigaci\'on. Pol\'\i gono la Coma.
    46980 Paterna, Valencia, Spain
\and
GEPI, Observatoire de Paris, CNRS, Universit\'e Paris Diderot; place Jules Janssen 92195 Meudon Cedex, France
%GEPI, UMR 8111 du CNRS, Observatoire de Paris-Meudon,  92195 Meudon, France
\and
GACE\--ICMUV
    Edicifio Institutos de Investigaci\'on. Pol\'\i gono la Coma.
    46980 Paterna, Valencia, Spain
\and
Instituto de Astrof\'\i sica de Andaluc\'\i a (CSIC). Camino Bajo de Hu\'etor, 24. 18008, Granada, Spain
}
 \authorrunning{J. Guti\'errez-Soto et al.}
 \titlerunning{A photometric study of Be stars located in the seismology fields of COROT}

  \date{Received ; accepted }

%opening

\abstract{%context
In preparation for the COROT mission, an exhaustive photometric study of Be stars located in the 
seismology fields of the COROT mission has been performed. The very precise and long-time-spanned photometric 
observations gathered by the COROT satellite will give important clues of the origin of the Be phenomenon.
}
{%aim
The aim of this work is to find short-period variable Be stars located in the seismology fields of COROT 
and to study and characterise their pulsational properties.
%The detection of a great number of variable Be stars has allowed us to study the pulsational properties of Be stars as a class.
}
{%methods
Light curves obtained at the Observatorio de Sierra Nevada together with data from Hipparcos and ASAS-3 of a total of 84 Be stars have been analysed in order to search 
for short-term variations.
We have applied standard Fourier techniques and non-linear least-square fitting to the time series. 
%A preliminary mode identification based on the photometric amplitude ratios has been applied to the periodic 
%Be stars.
}
{%results
We have found 7 multiperiodic, 21 mono-periodic and 26 non-variable Be stars. Short-term 
variability has been detected in 74\% of early-type Be stars and in 31\% of mid- to late-type Be stars. We have shown that non-radial pulsations are most frequent among Be stars than in slow-rotating B stars of the same spectral range. 
}
{%conclusions
%Be stars HD 49330, HD 50209, HD 50891, HD 51193, HD 51452, HD 175869 and HD 181231
%have high probability to be observed by the COROT satellite.
}

\maketitle

%---------------------------------------------------------------------------------%

\section{Introduction}

Be stars are non-supergiant B stars, that show or have shown 
at one or another moment emission in Balmer lines. It is 
generally agreed that the origin of this emission is the presence
of an equatorial circumstellar disk, fed by discrete mass loss events.
For a complete review of the ``Be phenomenon'' and its properties, see
\citet{porter03}.

Be stars show two different types of photometric variability, with different origin and 
time-scales:
(i) Long-term variability due to variations in the size and density of the circumstellar envelope. 
Variations are irregular and sometimes quasi-periodic, with time-scales of weeks to years. In some stars,
variations are in form of outbursts with total duration of weeks or months.
(ii) Short-term periodic variability, with time-scales from 0.1 to 2 days, 
generally attributed to non-radial pulsations.
\citet{hubert98}, based on Hipparcos observations, found that it is present in 86\% of early Be stars, 
in 40\% of intermediate sub-spectral types (B4e-B5e) and in only 18\% of late Be stars. 
In the HR diagram, early Be stars are located at the lower border of the instability domain of the $\beta$ 
Cephei stars, while mid and late Be stars are mixed with SPB stars. 
Both $\beta$ Cephei and SPB stars are main sequence pulsating B stars. Pulsations in $\beta$ Cephei stars
are caused by p-modes driven by the $\kappa$ mechanism associated with the Fe bump, and have periods
similar to the fundamental radial mode.
Slowly Pulsating B stars (SPB) are g-mode pulsators with periods longer than the fundamental 
radial one. Therefore, short-period p-modes are 
expected in early Be stars and long-period g-modes in mid to late Be stars.

Recently, the high-precision photometric data obtained with the MOST satellite has revealed a
rich spectrum of frequencies associated with radial and non-radial pulsations
in three Be stars, namely $\zeta$ Oph \citep[O9.5Ve, ][]{walker05b}, \object{HD 163868} \citep[B5Ve, ][]{walker05a} 
and \object{$\beta$ CMi} \citep[B8Ve, ][]{saio07}. 
This discovery suggests that pulsations are present in all 
rapidly rotating Be stars. 

The characterisation of the short-term variability in Be stars is essential for the
understanding of the Be phenomenon.
The spectroscopic analysis led by \citet{rivinius01} in $\mu$ Cen suggested that non-radial pulsations 
combined to the near break-up rotational velocity are 
probably the mechanism that cause the mass ejection.
However, $\mu$ Cen is, up to now, the only known Be star which presents this behaviour.
The observation of Be stars with the COROT satellite will provide 
photometric time series with such an unprecedented quality that will allow us to perform 
a deep study in the role of non-radial pulsations and its relation with the Be star outbursts.

The French led COROT satellite\footnote{http://corot.oamp.fr/} was successfully launched
on 27 December 2006 and started its scientific observations on 3 February 2007.
COROT has two main goals: the study of the stellar interiors by looking at their oscillations
and the search for exoplanets by detecting planetary transits \citep[see][]{baglin02}. 
In each pointing, up to 10 stars with magnitudes between 5.5 and 9.4 will be monitored in the two CCDs of the seismology program
during 150 or 20 days with a photometric precision of $\sim1$ ppm. 
The COROT telescope will be pointed alternatively
in the direction of two regions in the sky placed at the intersection 
between the equator and the galactic plane. Both regions have a radius 
of 10 degrees and are denoted as the centre and anticentre cones 
respectively, regarding their positions relative to the galactic center.

Bright Be stars will be therefore observed as secondary
targets of the asteroseismology program. An international collaboration 
(The COROT Be stars team\footnote{http://www.ster.kuleuven.ac.be/$\sim$coralie/corotbe.html})
led by A.-M. Hubert has studied the bright Be stars candidates to be observed by COROT. 
In this framework, \citet{neiner05b} identified 16 
previously unknown Be stars in the COROT observing cones, \citet{fremat06} 
determined the fundamental parameters of 64 Be stars
in the regions and
\citet{gutierrez07} found two of them as multiperiodic and
characterised their pulsational behaviour. 

The aim of this work is to characterise the short-term photometric variability of all bright Be stars located
in the COROT cones, which are suitable to be observed in the seismology fields.
In addition, we have investigated the pulsational properties of Be stars
%its relation with the classical B-type pulsators 
and 
their degree of variability with respect to
other pulsating B stars.

\section{Observations and data analysis}

\subsection{The sample}

A total of 84 Be stars have been found in the observing cones of COROT with magnitudes ranging from 5.5 to 9.4. 
The list has been taken from \citet{jaschek82}, complemented with the new Be stars identified by \citet{neiner05b}.

Due to the high number of stars to be analysed, a selection criterion has been made. 
As a first step, all the stars close to the primary targets of COROT have been observed 
during a 4-year campaign at the Observatorio de Sierra Nevada (OSN). 
They were firstly selected
as they have the highest probability to be observed by the satellite.
As the decision of the exact position of the CCDs has been finally taken 
by the scientific committee of COROT in 2006, we observed some Be stars which are not located
near the finally selected primary targets. 
Be stars whose results did not
lead to a convincing period determination during the observing run were re-observed 
at the OSN and re-analysed. 
In addition, the Hipparcos and ASAS-3 light curves of all Be stars in the COROT cones have been analysed in order to complement the study performed 
with the OSN data.

The studied stars with their spectral types and \vsini~are reported in Tables~\ref{table:summary:anti} and~\ref{table:summary:cen} (available only in electronic form at 
{\tt http://www.aanda.org}).

\subsection{Observatorio de Sierra Nevada (OSN)}
Ground-based observations were obtained at the 0.9 m telescope of the Observatorio de Sierra Nevada in Granada, Spain, 
from 2002 to 2006. 
The instrument used is the automatic four-channel spectrophotometer which allows simultaneous observations 
through the four $uvby$ filters of the Str\"omgren system. 
Obtaining photometric light curves in four filters at the same time
allows us to confirm or reject uncertain periods. Only frequencies detected simultaneously in the $vby$ filters 
are considered as certain, as the signal to noise for the $u$ filter is significantly lower.

\begin{table}
\centering
\caption{Summary of observing nights at the OSN. The mean accuracy of each dataset is also provide (see text for details).}
\begin{tabular}{cccccc}\\[0.2cm]
\hline
\hline
Dates & Observing & \multicolumn{4}{c}{Accuracy (mmag)}\\

      &      nights &   u  & v & b & y \\
\hline
20-29 May 2002 & 10    &10  &6  & 6 & 7   \\
8-14 Jan 2003 & 4      &9   & 4 & 3 & 4   \\
1-13 Jul 2003 & 9     &8   & 4 & 4 & 6   \\
30 Jan - 8 Feb 2004 &4 &10  & 5 & 5 & 6   \\
13 - 24 Jul 2005 & 6   &10  & 5 & 5 & 5   \\
16 - 30 Jan 2006 & 5   &7   & 4 & 4 & 5   \\
\hline
\end{tabular}
\label{table:nights}
\end{table}

Dates of the observing runs and observing nights are reported in Table~\ref{table:nights}. 
We have applied the three-star differential photometry, similar to the one described by \citet{lampens05}. 
A variable, a comparison, a check star and the appropriate sky background were measured successively 
(sky \-- var \-- comp \-- check \-- sky \-- var \ldots).
Usually we observed up to 8 target stars every night and we repeated the measurements for each night. 
Only in the winter 2003 we devoted each clear night to survey an individual star. 

Sky level and mean extinction coefficients have been obtained for each night.
An amount of 22 Be stars have been observed at the OSN in the anticentre and the centre directions. 
These Be stars with the corresponding comparison and check stars are 
presented in Tables~\ref{table:summary:anti} and~\ref{table:summary:cen} respectively.
A few comparison stars have been found to be low-amplitude variables, as for example, HD 171\,305 and HD 182\,786
(Guti\'errez-Soto et al. in preparation)

The mean accuracy of the differential photometry, 
measured as the standard deviation of the difference between the comparison and check values
for the whole campaign, is, 
in all cases, less than 10 mmag in $u$ and 7 mmag in $vby$, as shown in
Table~\ref{table:nights}.

\subsection{Hipparcos}

Hipparcos (High Precision Parallax COllecting Satellite, \citealt{perryman97}) was an astrometric mission of the
European Space Agency dedicated to the measurement of positions, parallaxes and proper motions 
of stars. As a by-product of the astrometric mission, stellar magnitudes were 
repeatedly measured for each star on numerous occasions throughout the mission, 
resulting in an enormous collection of light curves.
\citet{hubert98} showed that Hipparcos is a useful tool for the study of variability of bright Be stars.
A total of 62 stars located in the observing cones of COROT were observed by Hipparcos.
An average 
of 100 datapoints spanning 1000 \-- 1100 days are provided for each observed star. The standard error of a 
measurement ranges from 6 mmag for stars with magnitude 6 to about 17 mmag for stars with magnitude 9.

%%%%%%%%%%%%%%%%%%%%%%%%%%%%%%%%%%%%%%%%%ASAS!!!

\subsection{ASAS-3}
The All Sky Automated Survey \citep[ASAS,][]{pojmanski02} is a project whose final goal is photometric monitoring
of approximately $10^{7}$ stars brighter than 14 magnitude.
ASAS-3 is the third stage of the ASAS project, which has surveyed 
the whole of the southern and part of the northern sky.
We have analysed the light curve of 50 Be stars fainter than 7.8 
located in the observing cones of COROT. 
Stars brighter than 7.8 have not been considered as they most likely saturate the detector.
An average of 200 datapoints spanning over 1700 days are provided for each star.
Since five different magnitudes depending on the aperture radius are 
provided, we have selected the one which produces less error, usually the 4th and 5th aperture in our case.
Only measurements with quality grade A have been analysed.
Errors of the ASAS-3 photometry range from 10 to 25 mmag for the studied stars.

\subsection{Frequency analysis}

\begin{figure}
  \centering
  \includegraphics[width=9cm]{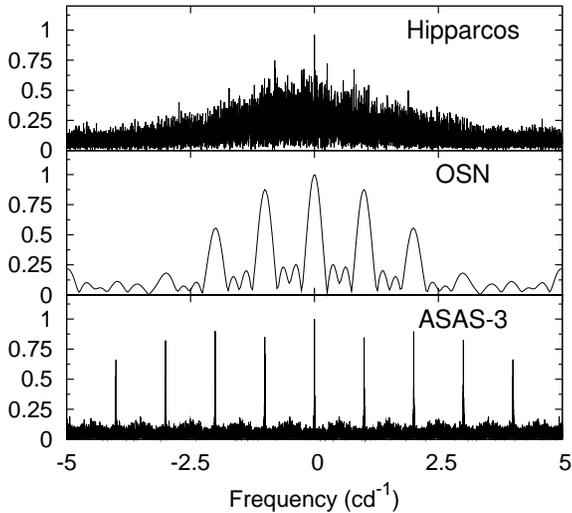}
  \caption{Examples of spectral window for the three instruments used in this paper.}
  \label{fig:window:1}
\end{figure}

Frequency analysis has been applied to the time series of each star in the three datasets
separately. In some cases we have been able to combine OSN \textbf{and} ASAS datasets in order to 
refine the detected frequencies.

We have employed the Period04 program \citep{lenz05}, which searches for frequencies by means of 
standard Fourier analysis.
Once a frequency is detected, the program adjusts the parameters of a sinusoidal function using a 
least-square fitting and prewhitens the signal from this frequency.
Then, a new step starts finding a new frequency, and the subsequent least-square fitting is performed 
allowing the two frequencies to move in order to get the minimum variance.
The method is iterative and stops when the removal of a new frequency is not 
statistically significant.

We have also used a non-linear multi-parameter fitting code which scans a 
wide range in frequency based on \citet{vanicek71} and explained in detail in 
\citet{zerbi97}. This code is well-suited for the OSN and ASAS-3 data, for which daily 
aliases are present in the periodogram, due to the fact that the observations 
have been obtained at only one site. We have to note here that the final frequency 
solution for the OSN and ASAS-3 data can be contaminated from 1 \cd\ aliases. 

In Fig.~\ref{fig:window:1} we show an example of spectral window for the Hipparcos, OSN and ASAS data.
The non-random distribution in time of the Hipparcos photometry produces a spectral window
which is very complex. In the case of the OSN and ASAS-3 observations, a 1-day alias pattern 
is present in the spectral window. Note the different widths of the sidelobes, depending on the time length 
of the observations.

We accept frequencies as long as they fulfill the signal-to-noise ratio (SNR) criterion, described in 
\citet{breger93}. \citet{kuschnig97} demonstrated that a level in the amplitude spectrum of 4 times the noise level will 
include the 99.9\% of all peaks due to noise.  
The noise level is computed by averaging the amplitude within a 5 \cd\ frequency interval 
of the residual periodogram after final prewhitening.

For the determination of the error in frequencies, we follow equations given by \citet{montgomery99}, 
taking into account the correlations in the residuals of the fitting \citep{czerny91}. 
We have obtained an error in frequency of $3\--4 \times 10^{-5}$, $2\--4 \times 10^{-3}$ and $1\--2 \times 10^{-5}$ for 
the Hipparcos, OSN and ASAS-3 datasets respectively. 
The frequency resolution is given by $1/\mathrm{T}$, where T is the time length of the observations. In our case, 
we estimate 0.1 \cd, 0.001 \cd\ and 0.0006 \cd\ for the OSN, Hipparcos and 
ASAS-3 observations respectively.

\subsection{Results}

Results of the data analysis are summarised in Tables~\ref{table:resultados:anti} and~\ref{table:resultados:cen},
where we present the frequencies found for all stars analysed for short-term variability.
In the few cases in which we have obtained different results from different datasets, we considered
as more reliable the one from the instrument with higher precision. 
Therefore, frequencies obtained from the OSN data are considered as the most reliable, 
followed by those found from the Hipparcos and ASAS-3 datasets. 

Mid- and long-term variability is often present in the light curves of Be stars.
In some cases, it shows a complex pattern which prevents the study of the short-term
variability.
In addition, the data for some stars present a bad sampling or too few datapoints in the 
light curve.
For these reasons, we have not been able to perform short-period analysis for some stars
in our sample. They are marked  
with the ``-'' symbol in Tables~\ref{table:resultados:anti} and~\ref{table:resultados:cen}.

Stars showing short-term variability with peak to peak amplitude higher than 0.04 magnitudes during several days
are considered as variable, even if we have not been able to determine the frequency.
For other stars, the analysis performed in this work yields to uncertain frequencies.  
These stars are marked in Tables~\ref{table:resultados:anti} and~\ref{table:resultados:cen} with the symbols 
 ``var'' and ``?'' respectively. 

Short-term variability analysis has been finally performed on 57 bright Be stars in the COROT cones. 
31 have been found to be variable and 26 non-variable at the detection level of the instruments. 
Multiperiodic variability has been found in seven stars. 
Notes on the individual stars are presented in Sect.~\ref{sect:notes}, 
only available in the electronic edition of the paper,
which includes Tables~\ref{table:summary:anti} and~\ref{table:summary:cen} and Figs.~\ref{fig:42259:1} to~\ref{fig:183656:1}.

\section{Discussion}

%---emp tabla larga
\begin{table*}
\centering
\caption{Results of the photometric study for stars in the Galactic Anticentre direction.
Frequencies in \cd\ are shown for the Hipparcos, OSN and ASAS-3 datasets. The '\--' 
stands for stars for which a short-period analysis could not be performed. 
The '*' stands for stars that have not been observed 
with the instrument. 
The '**' stands for stars that are saturated in the ASAS-3 database.
The 'var' stands for stars showing short-term variability but not periodicity. 
The 'no var' stands for stars that do not show short-term variations at the detection level of the instrument.
The '?' stands for frequencies which are uncertain. For details, see text. 
\textbf{In the last column some complementary indications and/or frequencies detected by other authors are given:}
(1) - Star included in the ASAS-3 catalogue of variable stars \citep{pojmanski02};
(2) - \citet{neiner05b}; (3) - \citet{hubert98}; 
(4) - \citet{gutierrez07}; (5) - \citet{lynds60};
(6) - \citet{percy02}.
}
\begin{tabular}{ccccc}
%{@{\ }c@{\ \ \ }c@{\ \ \ }c@{\ \ \ }c@{\ \ \ }c@{\ \ \ }}

\hline
\hline
 Be star & Hipparcos & OSN & ASAS-3 & Remarks\\
\hline
\object{HD 42259}   &   var     & 1.28        & 1.3223 + 0.7401 & \\
\object{HD 42406}   &   -      & *           & no var          & \\
\object{HD 43264}  &   -     & *           & **              & \\ %doble
\object{HD 43285}   & 2.203/6.173?   & no var      & **              & \\
\object{HD 43777} &  -         & *           & no var          & \\%7.8
\object{HD 43913} &  -         & *           & no var          & \\%doble
\object{HD 44783} &    -       & *           & **              & \\
\object{HD 45260}   &   var       & *           & -               &\\% long-term\\
\object{HD 45626}   &    *     & *           & -               & \\
\object{HD 45901}   &   0.466       & *           & -  & \\%long-term\\ %anyadida
\object{HD 45910} &    -       & *           & **              & \\ %doble
\object{HD 46380}   &   -     &1.75  ?      & -              & \\%largo 
\object{HD 46484}   &    -     &2.45         & **              & \\
\object{HD 47054} &   no var        & *           & **              & \\
\object{HD 47160} &   -        & *           & **              & \\% doble
\object{HD 47359}   &    *     & *           & -              & \\
\object{HD 47761}   &    -       & *           & -              & \\%outbursts\\%, long-term \\ %hipp tb
\object{HD 48282}   &    -     &   *         & 0.6819          & \\%long-term\\ %long-term hipp + ASAS
\object{HD 49330}   & 3.534    &2.129        & -               & (1)\\ %long-term
\object{HD 49567} &   0.39       & *           & **              & 0.39 (2)\\% ;0.40 (5)\\ 
\object{HD 49585}   &    *     &1.65 + 1.21  & -               & (1)\\ %long-term
\object{HD 49787} &  -         & *           & **              & \\
\object{HD 49992}   &    *     & *           & var             & \\%no period
\object{HD 50083} &   -        & *           & **              & \\%long-term\\ %herbig en ref
\object{HD 50087}   &  -       &     no var  &no var          & \\
%---------HD 50138 &   ref B[e] u        & *           & **              & \\ %NO LA PONGO
\object{HD 50209}   & 1.689/1.47  &1.4889       &2.4803+1.4749    & \\ 
\object{HD 50424}   &    *     & *           & -               &\\%long-term\\
\object{HD 50581} &  no var         & *           & **              & \\% 
\object{HD 50696}   &    *     &3.22         & -                & \\%long-term \\ %largo
\object{HD 50820} &   no var        & *           & **              & \\% doble
\object{HD 50868} &   -         & *           & no var             & \\%long-term\\% long-term hipp, outbursts
\object{HD 50891}   &    *     &1.88         & -               & (1)\\ %largo
\object{HD 51193}   &1.639     &2.66         & 1.606           & \\
\object{HD 51404}   &    *     &2.68 + 5.99  & var             & \\ %no period in ASAS
\object{HD 51452}   &    *     & 1.58 ?        & no var          & \\
\object{HD 51506} &  -         & *           & **              & \\
%---------HD 52721 &  posible herbig doble var         & *           & **              & \\% NO LA PONGO
\object{HD 53085} &  -         & *           & **              & \\%long-term\\
\object{HD 53367} &  -         & *           & **              & \\%long-term\\
%---------HD 53667 &  herbig         & *           & **              & \\ %NO LA PONGO
\object{HD 54464}   &    *     & *           & -               & \\%long-term\\
\object{HD 55135} &  -         & *           & **              & \\
\object{HD 55439}   &   -       & *           &  -             & \\%outbursts\\
\object{HD 55606}   &    *     & *           & no var              & \\
\object{HD 55806}   &    -      & *           & no var              & \\%long-term\\ %hipp
\object{HD 57386}   &   -      & *           & no var              & \\ %doble
\object{HD 57539} & -          & *           & **              & \\%long-term \\
\object{HD 259431}  &   -       & *           & no var              & \\
\object{HD 259440}  &    *     & *           & no var              & \\
\object{HD 259597}  &    *     & *           & no var              & \\
\object{HD 293396}  &    *     & *           & -              & \\%long-term\\
\hline
\end{tabular}
\label{table:resultados:anti}
\end{table*}

%fin tabla result-------------

%%%%%%%tabla con todas las estrellas!! COMPROBAR, HACER HIPPARCOS!!
%---emp tabla result

\begin{table*}
\centering
\caption{Same as for Table~\ref{table:resultados:anti}, but for stars in the Galactic Centre  direction.}
\begin{tabular}{ccccc}
\hline
\hline
 Be star & Hipparcos & OSN & ASAS-3 & Remarks\\
\hline
\object{HD 166917} & no var        &*         & **                   & \\
\object{HD 168797} & 2.049          & 1.20+1.13+3.30+1.41.    &**                 & 2.049 (3); 1.20+1.13+3.30+1.41 (4)\\
\object{HD 170009} & no var       &*         & *                   & \\
\object{HD 170714} &   -    &3.79         & **              & \\
\object{HD 171219} & no var  &no var       & **              & \\ 
\object{HD 173219} &   var     &*         & **                   & \\
\object{HD 173292} &   var     &   *         &0.6824           & \\
\object{HD 173371} &   no var       &*         & **                   & \\
\object{HD 173530} &  *      &   *         &1.3703           &  \\
\object{HD 173637} &  -      &1.86         &-                & (1)\\ %long-term
\object{HD 173817} &  no var      &3.51 ?       & no var          & \\ 
\object{HD 174512} &  0.82      &   *         & no var          &0.82 (2)\\ 
\object{HD 174513} & -     &3.34       & 0.0271 + 5.293  &\\%long-term \\%long-term hipp
\object{HD 174571} &  -      &   *         & -               &\\ %doble
\object{HD 174705} &  *      &   *         & 2.3164          &\\
\object{HD 174886} &   no var      &*         & **                   & \\
\object{HD 175869} &  no var      &*         & **                   & \\
\object{HD 176159} &  -      &*         & no var                  & \\%long-term\\ %hipp
\object{HD 176630} &     1.59     &*         & **                   & 1.59 (2)\\
\object{HD 178479} &  -      &*         & -                   &\\%long-term \\ %tb hipp
\object{HD 179343} &  -      &*         & **         & \\ %doble
\object{HD 179405} &   var         & 1.62+2.78+2.56+1.27    & -                 & 1.62+2.78+2.56+1.27 (4)\\
\object{HD 180126} &   -       &*         & -                   & \\%outbursts\\%, long-term \\ %tb hipp
\object{HD 181231} &  no var     &  0.67?      & 3.4304?          & \\ 
\object{HD 181308} &  *      &*         & -                   & \\
\object{HD 181367} &  *      & no var      & no var               & \\  
\object{HD 181709} &  no var       &*         & no var                   & \\
\object{HD 181803} &   -      &*         & -                  & \\
\object{HD 183656} &  1.534  &  3.63 & **              & 1.534 (3);1.1739  (5)\\
\object{HD 184203} &  *      &*         & -                   & \\
\object{HD 184279} &  6.410 ?& var &  **             &  (1); 1.667  (6)\\
\object{HD 184767} &   -     &*         & **                   & \\% doble
\object{HD 194244} &  no var       &*         & **                   & \\
\object{HD 230579} &  *       &*         & no var                   & \\
\object{BD-094858} &  *       &*         & -                    &  \\
\hline
\end{tabular}
\label{table:resultados:cen}
\end{table*}
%fin tabla result-------------

\begin{figure}
  \centering
  \includegraphics[width=8cm]{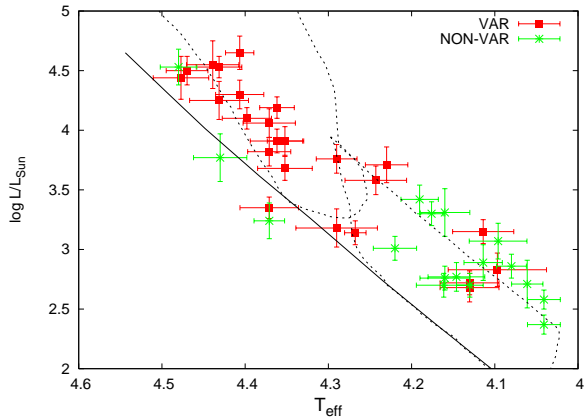}
  \caption{Location of the studied Be stars in the theoretical HR diagram.
Red \textbf{filled squares} correspond to variable stars and green \textbf{asterisks} to non-variable stars. 
The solid line represents the ZAMS from \citet{schaller92} and
the dashed lines describe the theoretical $\beta$ Cephei and SPB instability strips 
computed by \citet{pamyatnykh99}. 
}
  \label{fig:hr:1}
\end{figure}

In Fig.~\ref{fig:hr:1} we show the position in the HR diagram of the stars
for which we have performed short-term variability analysis.
Only 41 Be stars of this sample have accurate determination of 
their physical parameters.
Values of $\log \mathrm{L}/\mathrm{L}_{\sun}$ and \teff~of all stars but one
are taken from \citet{fremat06}, assuming \omc = 0.9, which is the average 
angular velocity rate of galactic field Be stars \citep{fremat05}.
For the star HD 48282, the spectral parameters are taken from the
paper by \citet{levenhagen04}.
For comparison, we have also plotted the theoretical boundaries of the $\beta$ Cephei and SPB instability strips 
from \citet{pamyatnykh99}.

We investigate the degree of short-term variability
in our sample of Be stars. 
We have distinguished between early (B0-B3) and mid- to late-type Be stars (B4-B9).
In the case of pulsating B-type stars, the first interval is occupied by $\beta$ Cephei variables and 
the hottest SPB stars. According to \citet{pamyatnykh99}, the hot-temperature boundary 
computed with OPAL opacities takes place at B3 (see Fig.~\ref{fig:hr:1}), but if OP opacities are used, 
g-modes are excited even in stars much hotter than B3. 
Pulsating B stars in the B4-B9 interval are only SPB stars. 
In addition, this distinction between early- and late-type Be stars has also been
considered by other authors  \citep[eg.][]{hubert98} from phenomenological reasons: 
pulsations are frequent in the early-type interval and scarce in the late-type one. 

\begin{figure}
  \centering
  \includegraphics[width=8cm]{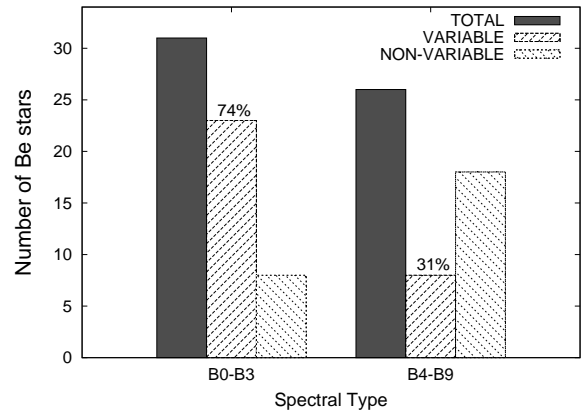}
  \caption{Distribution of short-term variability as a function of spectral type for stars in our sample.}
  \label{fig:estad:1}
\end{figure}

We have found that short-term variability is present in 74\% of early-type Be stars 
and 31\% of mid- to late-type Be stars (see Fig.~\ref{fig:estad:1}). 
The results presented here are similar to those obtained by \citet{hubert98},
from a larger sample of Be stars observed by the space mission Hipparcos.

In order to compare the degree of variability of Be stars with
respect to slow-rotating B-type stars, we have searched in the literature for 
the fraction of $\beta$ Cephei and SPB stars among the total number of B stars in 
their spectral ranges. To our knowledge, general studies of these fractions 
have not been undertaken. 
For this reason, we have studied the fraction of $\beta$ Cephei and SPB stars in a
sample of galactic field stars observed by the Hipparcos mission. 
Our sample consists of all stars in the Bright Star Catalogue  \citep{hoffleit91} with 
spectral type between B0 and B9 and luminosity class V-III, whose 
Hipparcos parallax has an accuracy better than 20\% and with 
Str\"omgren photometry in the catalogue of \citet{hauck99}. 
All known Be stars have been excluded from the sample.
The whole sample consists of 185 stars in the B0-B3 spectral domain
and 610 in the B4-B9 domain. The Hipparcos photometry of all these stars
has been searched for short-term variability \citep{eyer98,waelkens98}.
The results of these surveys together with previous literature 
are included in the most recent lists of known $\beta$ Cephei \citep{stankov05}
and SPB stars \citep{decat02}.

Among the 185 stars in the early spectral domain, 16 are catalogued as
$\beta$ Cephei and 14 as SPB stars, considering both confirmed and candidate SPB stars as listed by \citet{decat02}.
This amounts to a total of 30 pulsating stars, representing the 16\% of the sample. 
For stars in the B4-B9 range, 21 are listed as confirmed and candidate SPB stars. 
This represents a fraction of 3.4\%.
The photometric data and the variability search techniques for this 
sample are the same as \citet{hubert98} used for their 
short-term variability study of Be stars observed by Hipparcos, and 
hence the results of both searches are directly comparable. 

In addition, we have searched in the literature for photometric surveys
aimed to detect pulsating B stars in the population of galactic
open clusters which are young enough to have their B-star main sequence complete. 
A cluster population cannot be considered as 
representative of the galactic field as it is composed by stars in a very narrow 
range of ages and metallicities. However, by taking into account the mean results
for several clusters with different astrophysical parameters, we can consider
them representative.

The number of $\beta$ Cephei and SPB stars in clusters have been taken from the studies 
of \citet{krzesinski97} for NGC 884, \citet{krzesinski99} for NGC 869, 
\citet{pigulski01} for NGC 663, \citet{balona94_a} for NGC 3293, 
\citet{balona94_b} for NGC 4755, \citet{balona95_b} and \citet{arentoft01} for 
NGC 6231 and \citet{balona96} for NGC 2362. 
In order to select the B stars in the B0-B3 and B4-B9 spectral ranges, 
we have used Str\"omgren $uvby$ photometry from \citet{capilla02} for 
NGC 884 and NGC 869, \citet{fabregat05} for NGC 663, \citet{balona94_a} for NGC 3293, 
\citet{balona94_b} for NGC 4755, \citet{balona95_b} for 
NGC 6231 and \citet{balona96} for NGC 2362.

Stars within the two ranges have been selected by using the
reddening-free indices [c1] and [m1] and their relationship with spectral types given by \citet{moon86}.
Known Be stars have been removed from the final list.

In Table~\ref{table:estad:cumulos} we show the percentage of $\beta$ Cephei in 
the studied open clusters. An average percentage of around 13.5\% is found for $\beta$ Cep stars. Note that 
the dispersion is quite high, although it is already known that clusters with similar age 
have very different proportion of $\beta$ Cephei stars \citep{balona97}.
The number of SPB stars found in the above clusters is too low to allow 
any reliable statistics.

The fraction of early-type pulsating B stars is 16\% and 13.5\% from the galactic field and cluster stars respectively.
These values are significantly lower than the percentage of 
pulsating Be stars with spectral types between B0 and B3, which is 74\%
from the present work, and 86\% from the Hipparcos sample studied by \citet{hubert98}.
The fraction of SPB stars in our late-type B stars sample is 
3.4\%, again significanly lower than the value of 
31\% found for late-type Be stars in our sample and
26\% found by \citet{hubert98}. We can conclude that 
the fraction of non-radial pulsators among fast-rotating Be stars
is much higher than the fraction among slow-rotating B stars
of the same spectral type.

One of the most remarkable characteristic of Be stars is the 
high rotational velocity. In this context, the above results 
suggest that high-rotational velocity may have the effect to trigger the
development of non-radial pulsations in B stars, or to enhance the 
amplitude of existing modes to make them more easily detectable. 
Alternatively, the higher prevalence of non-radial pulsations could 
be related to other ingredient of the yet unknown nature of the Be 
phenomenon.

A deeper exploration of the physical implications of these results is not possible yet,
due to the lack of adequate models describing the stellar pulsations at rotational 
velocities close to the critical. However, we consider that these results constitute
a valuable input for the models being currently developed \citep[e.g. ][]{reese06}.

\begin{table}
\centering
\caption{Percentage of $\beta$ Cephei stars among the number of B stars in the same spectral range for different open clusters.}
\begin{tabular}{cc}
\hline
\hline
 Cluster & \% $\beta$ Cep \\
\hline
NGC 884   & 12-14\%  \\
NGC 869   &  5-7\%   \\ 
NGC 663   &  9-12\%  \\
NGC 3293  &  25\%    \\
NGC 4755  &  25-30\% \\
NGC 6231  &  10-15\% \\
NGC 2362  &   0\%    \\
\end{tabular}
\label{table:estad:cumulos}
\end{table}
%fin tabla result-------------

\section{Conclusions}

We have studied the short-term variability of 57 bright Be stars located in the seismology fields of COROT. 
We have analysed $uvby$ photometry obtained at the OSN together with
data from the Hipparcos mission and ASAS-3 project.
31 stars have been 
detected as variable and 26 are considered as non-variable
at the detection level of the instruments. 
Moreover, seven variable stars have been found 
to be multiperiodic. 

We have shown that non-radial pulsations are most 
frequent \textbf{in} Be stars than in slow-rotating B stars 
of the same spectral range. These results allow us to suggest that
high-rotational velocity can either contribute to trigger the 
development of non-radial pulsations or to enhance the 
amplitude of the existing modes.
As an alternative explanation, the prevalence of non-radial
pulsations could be related to the yet unknown nature of the Be
phenomenon.

The observations of Be stars with the COROT satellite, 
currently underway, will give an answer to these
issues and serve as input to the elaboration of pulsational models for 
high-rotating stars which are currently being developed.

\begin{acknowledgements}
We would like to thank Ennio Poretti for allowing us to use his data from San Pedro Martir and
Katrien Uytterhoeven for her useful comments.
This research is based on data obtained at the Observatorio de Sierra Nevada, which is 
operated by the CSIC through the Instituto de Astrof\'\i sica de Andaluc\' \i a.
%The work of J.G-S. is supported by a FPU grant from the Spanish ``Ministerio de Educacion y Ciencia''.
J.F. and J.S. acknowledge financial support from the program ESP~2004-03855-C03.

\end{acknowledgements}

\bibliographystyle{aa}
\bibliography{8252}

%%%%%%%%%%%%%%%%%%%%%%%%%%%%%%%%%%%%%%%%%%%%%%%%%%%%%
\Online

\begin{table*}
\centering
\caption{Studied Be stars in the Galactic Anticentre direction. 
ID numbers and SIMBAD V magnitudes are given for each target in cols.~1 and 2.
The spectral type and \vsini~are gathered in cols.~3 and 4. 
Comparison and check stars used for the differential 
photometry are given in cols.~5 and 6. References of the spectral type and \vsini~are given in col.~7.
References: : (1) \-- \citet{fremat06}; 
(2) \-- Simbad database. }
\begin{tabular}{ccccccc}
%{@{\ }c@{\ \ \ }c@{\ \ \ }c@{\ \ \ }c@{\ \ \ }c@{\ \ \ }c@{\ \ \ }c@{\ \ \ }}
\hline
\hline
Be star       & V   &Sp.Type  & \vsini   &   Comp.       & Check & References\\
              &     &         &  \kms  &               &       & \\
\hline
HD 42259 &8.89  & B0V    &          & \object{HD 294788}    & \object{HD 42369}  &  2\\
HD 42406 & 8.0 & B4IV    & 300          &                      &                     & 1 \\ 
HD 43264 & 6.05 & B9III    & 284          &                      &                     & 1 \\
HD 43285 &6.05  & B5IV   & 274          & \object{HD 44783}    & \object{HD 43526}   & 1\\
HD 43777 & 7.8 & B5    &               &                      &                     & 2 \\ 
HD 43913 & 7.88 & A0    &               &                      &                     & 2 \\ 
HD 44783 & 6.23 & B9II    & 227          &                      &                     & 1 \\
HD 45260 & 9.04 & B8    &               &                      &                     & 2 \\ 
HD 45626 & 9.25 & B7    &               &                      &                     & 2 \\ 
HD 45901 & 8.85 & B0.5IV &173           &                      &                     & 1 \\ 
HD 45910 & 6.74 & B2III    &           &                      &                     & 2 \\
HD 46380 &8.05  & B1.5IV & 310          & \object{HD 46541}    &  \object{HD 46519}  & 1\\
HD 46484 &7.65  & B0.5IV & 130          & \object{HD 46106}    & \object{HD 46748}   & 1 \\ 
HD 47054 & 5.52 & B7III  & 229              &                      &                     & 1 \\ 
HD 47160 & 7.10 & B8IV    & 158          &                      &                     & 1 \\
HD 47359 & 8.87 & B0IV    & 469          &                      &                     & 1 \\ 
HD 47761 & 8.72 & B2V    & 50          &                      &                     & 2 \\ 
HD 48282 & 8.79 & B3V    & 188          &                      &                     & 2 \\ 
HD 49330 &8.95 & B0.5IV  & 285          & \object{HD 50086}    &  \object{HD 50230}  & 1\\
HD 49567 & 6.15 & B3III  & 94              &                      &                     & 1 \\ 
HD 49585 &9.13 & B0.5IV  & 325          & HD 50086    &  HD 50230   &1\\
HD 49787 & 7.55 & B1V    & 169          &                      &                     & 1 \\
HD 49992 & 8.98 & B1     &           &                      &                     & 2 \\ 
HD 50083 & 6.91 & B2III    & 193          &                      &                     & 1 \\
HD 50087 &9.08  & B8III  &         & HD 50086    &  HD 50230   &2\\
HD 50209 &8.36 & B8IV    & 209          & HD 50086    &  HD 50230   &1\\
HD 50424 & 8.92 & B9    &           &                      &                     & 2 \\ 
HD 50581 & 7.54 & A0IV  & 250              &                      &                     & 1 \\ 
HD 50696 &8.87 & B1.5III & 366          & HD 50086    &  HD 50230   &1\\
HD 50820 & 6.27 & B3IV    &               &                      &                     & 2 \\ 
HD 50868 & 7.87 & B1.5V  & 276              &                      &                     & 1 \\ 
HD 50891 &8.88 & B0.5V   & 231          & \object{HD 50348}    & \object{HD 51150}    &1 \\
HD 51193 &8.06 & B1.5IV  & 224          & HD 50348    & HD 51150     &1\\
HD 51404 &9.30 & B1.5V   & 353          & HD 50348    & HD 51150     &1\\
HD 51452 &8.08 & B0IV    & 309          & HD 50348    & HD 51150    &1\\
HD 51506 & 7.68 & B2.5IV    &186               &                      &                     & 1 \\
HD 53085 & 7.21 & B4IV    & 222          &                      &                     & 1 \\ 
HD 53367 & 6.97 & B0IV    & 70          &                      &                     & 2 \\ 
HD 54464 & 8.4 & B2.5III    &177               &                      &                     & 1 \\
HD 55135 & 7.34 & B2.5V    &264               &                      &                     & 1 \\
HD 55439 & 8.47 & B2    &               &                      &                     & 2 \\
HD 55606 & 9.04 & B0.5V    & 361          &                      &                     & 1 \\ 
HD 55806 & 9.11 & B7III    & 202          &                      &                     & 1 \\ 
HD 57386 & 8.0 & B1.5V    &           &                      &                     & 2 \\ 
HD 57539 & 6.6 & B5III    &155               &                      &                     & 1 \\
HD 259431 & 8.71 & B6    & 95          &                      &                     & 2 \\ 
HD 259440 & 9.12 & B0    & 430          &                      &                     & 2 \\ 
HD 259597 & 9.33 & B1V    &           &                      &                     & 2 \\ 
HD 293396 & 8.59 & B1V    &           &                      &                     & 2 \\ 
\hline
%\multicolumn{7}{l}{References: (1) \-- \citet{fremat06}; 
%(2) \-- Simbad database.}
\end{tabular}
\label{table:summary:anti}
\end{table*}

%----------------fin tabla -------------

%-------------emp tabla------------
\begin{table*}
\centering
\caption{The same as for Table~\ref{table:summary:anti}, but for the stars in the 
Galactic Centre direction.}
\begin{tabular}{ccccccc}
\hline
\hline
Be star       & V   &Sp.Type  & \vsini &   Comp.  & Check & References\\
&     &         &  \kms  &               &       & \\
\hline
HD 166917 & 6.69& B8III      & 173             &                        &                    &1 \\  
HD 168797 & 6.14 & B2.5III   & 279     &  \object{HD 170200} & \object{SAO 123607}   &1  \\
HD 170009 & 8.00& B2.5III      & 181             &                        &                    &1 \\  
HD 170714 & 7.38& B1.5IV  & 280          &   \object{HD 171305}   & \object{HD 169581} &1 \\ 
HD 171219 & 7.65& B5III   & 314          &   HD 170200   & SAO 123607&1 \\
HD 173219 & 7.82& B0.5IV  & 66             &                        &                    &1 \\  
HD 173292 & 8.60& B8      &              &                        &                    &2 \\  
HD 173371 & 6.89& B7IV      & 295             &                        &                    &1 \\  
HD 173530 & 8.87& B7III   & 246         &                        &                   &1\\ 
HD 173637 & 9.29& B1IV    & 207          &   \object{HD 173693}   &\object{HD 173850} &1\\
HD 173817 & 8.65& B6IV    & 276          &   \object{HD 172868}   &\object{HD 173422} &1\\
HD 174512 & 8   & B8      &              &                        &                    &2 \\  
HD 174513 & 8.70& B1.5IV  & 261          &   \object{HD 174395}   &\object{HD 174650} &1 \\
HD 174571 & 8.89& B1.5V   & 250             &                        &                    &1 \\  
HD 174705 & 8.34& B1.5IV  & 341          &                        &                   &1 \\    
HD 174886 & 7.77& B4III      & 79             &                        &                    &1 \\  
HD 175869 & 5.56& B8III      & 171             &                        &                    &1 \\  
HD 176159 & 8.98& B5IV      & 243             &                        &                    &1 \\  
HD 176630 & 7.70& B3III      & 188             &                        &                    &1 \\  
HD 178479 & 8.92& B3V      & 109             &                        &                    &1 \\  
HD 179343 & 6.94 & B8III    &155               &                      &                     & 1 \\
HD 179405 & 9.12 & B2IV    &  248               &\object{HD 179846} & \object{HD 178598} &1\\
HD 180126 & 8.00& B2IV      & 252             &                        &                    &1 \\  
HD 181231 & 8.58& B5IV    & 259          &   \object{HD 182198}   &\object{HD 182786} &1  \\
HD 181308 & 8.70& B5IV      & 261             &                        &                    &1 \\  
HD 181367 & 9.36& B6IV    & 292          &   HD 182198   &HD 182786   &1\\
HD 181709 & 8.79& B6III      & 291             &                        &                    &1 \\  
HD 181803 & 9.10& B7III      & 190            &                        &                    &1 \\  
HD 183656 & 6.09& B6V & 270          &   \object{HD 183227}   &\object{HD 183563} &2\\
HD 184203 & 9.16& B9      &             &                        &                    &2 \\  
HD 184279 & 6.98& B0V     & 137          &   HD 183227   &HD 183563   &1\\
HD 184767 & 7.18 & A0III    &49               &                      &                     & 1 \\
HD 194244 & 6.14& B9III      & 232             &                        &                    &1 \\  
HD 230579 & 9.10& B1IV      & 343            &                        &                    &1 \\  
BD\,-094858 & 8.84& B1.5V      & 115            &                        &                    &1 \\  
\hline
\end{tabular}
\label{table:summary:cen}
\end{table*}
%----------------fin tabla -------------

\section{Notes on individual stars}\label{sect:notes}

Here we present some notes on all the Be stars observed at the OSN and the Be
stars that have shown periodicity in the ASAS-3 or Hipparcos light curve.
For a detail study of stars HD 168797 (NW~Ser) and HD 179405 (V1446~Aql), see \citet{gutierrez07}.
We have divided the sample in stars located in the Galactic Centre and Anticentre 
directions.

\subsection{Stars in the Galactic Anticentre direction}

%%%%%%%%%%%%%%%%%%%%%%%%%%%%%%%%%%%%%%%%%%%%%%%%%
\subsubsection{HD 42259} 

The Hipparcos light curve shows variability with 
an amplitude from peak to peak of 6 hundredths of magnitude. 

\begin{figure}
  \centering
  \includegraphics[width=8cm]{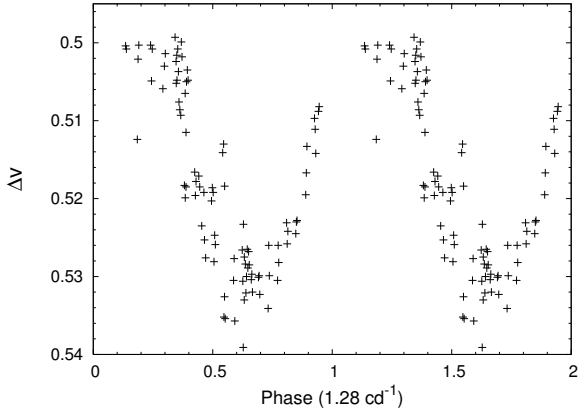}
  \caption{Light curve of the star HD 42259 folded in phase with the frequency $1.28$ \cd, for the OSN data in the $v$ filter.}
  \label{fig:42259:1}
\end{figure}

We observed this star in 2006 at the OSN.
The amplitude of the light curve varies from 
night to night, suggesting the presence of multiperiodicity.
Strong peaks appear in the periodogram of the 
$v$ filter at frequency $1.28$ \cd\ and its 1-day aliases. Results with the $by$ filters are similar
within errors. A phase plot with this frequency is displayed in Fig.~\ref{fig:42259:1}.
A frequency at $1.41$ \cd\ has also a good fit and good phase diagram. The 
time span of the observations does not allow us to distinguish between these two frequencies. 

\begin{figure}
  \centering
  \includegraphics[width=9cm]{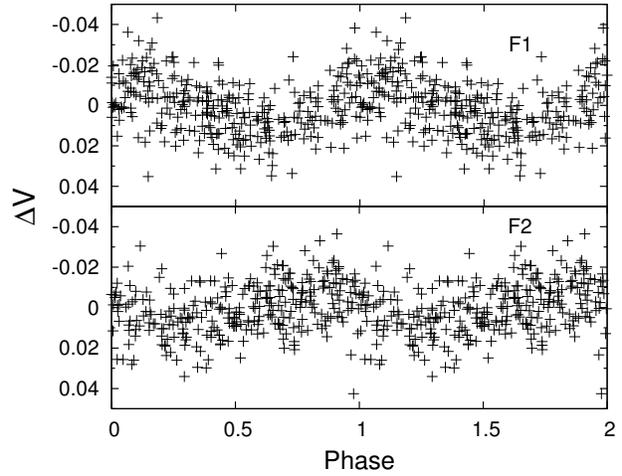}
  \caption{Light curve of the star HD 42259 folded in phase with the frequency $\mathrm{F}1=1.3223$ \cd\ 
 for the ASAS-3 dataset (\textbf{top})
 and with the frequency $\mathrm{F}2=0.7401$ \cd\ for the residuals after prewhitening for F1 (\textbf{bottom}).}
  \label{fig:42259:2}
\end{figure}

In the analysis of the ASAS-3 dataset, we find a frequency at $\mathrm{F}1=1.3223$ \cd (Fig.~\ref{fig:42259:2}, upper panel), which is 
similar within errors to the one detected at the OSN. 
After prewhitening for this frequency, we  
find another significant frequency at $\mathrm{F}2=0.7401$ \cd. The 1-day alias $\mathrm{F}2'=1.2626$ \cd\ gives also a similar fit, although 
F2 has a less scattered phase diagram (Fig.~\ref{fig:42259:2}, bottom panel).

%%%%%%%%%%%%%%%%%%%%%%%%%%%%%%%%%%%%%%%%%%%%%%%%% TERMINADO
\subsubsection{HD 43285} 
Peaks at frequencies $2.203$ and 6.173 \cd\ are detected in the Hipparcos data,
although the 
resulting phase diagrams are very scattered.

We observed this star during only one season in 2003 at the OSN.
Our light curve spanning 9 hours 
does not show any sign of variation at any frequency with an amplitude
exceeding 2 mmag in the $vby$ filters and 3 mmag in the $u$ filter.
This star is saturated in the ASAS-3 database.

%%%%%%%%%%%%%%%%%%%%%%%%%%%%%%%%%%%%%%%%%%%%%%%%% TERMINADO
\subsubsection{HD 45901} 
\begin{figure}
  \centering
  \includegraphics[width=8cm]{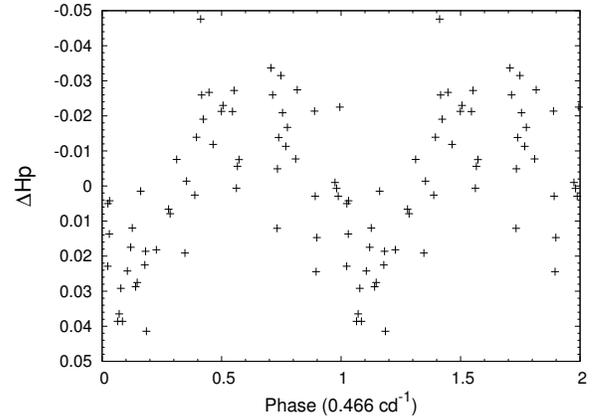}
  \caption{Light curve of the star HD 45901 folded in phase with the frequency $0.466$ \cd\ 
 for the Hipparcos dataset. }
  \label{fig:45901:1}
\end{figure}

This star has only been observed with Hipparcos and ASAS-3. 
A long-term trend is present in the Hipparcos light curve.
After prewhitening for this trend, significant peaks at frequencies 
0.548 \cd\ and 0.466 \cd\ are clearly detected in the periodogram.
The less scattered phase diagram is obtained with frequency 0.466 \cd, which is shown in 
Fig.~\ref{fig:45901:1}.

The ASAS-3 data show a long-term variation complicated to be modelled 
which prevents us to search for short-term variability
in this dataset.

\subsubsection{HD 46380} 
The Hipparcos light curve shows mid- and long-term variability, which does not 
allow us to search for short-term periodicity.

The photometric observations collected at the OSN extend over two seasons, 2003 and 2006.
The 2003 dataset confirms the variability of the star.
A frequency of 3.49 \cd\ appears clearly in the periodogram, although it is very uncertain, due to the 
short time coverage (6 hours). 
The light curve obtained in 2006 shows clear variability at frequency $1.75$ \cd\ in all filters. 
The phase diagram for this frequency is very scattered, 
due to the fact that the amplitude of the signal is 3 mmag.
Note that the frequency obtained in 2003 is twice the frequency found in 2006. The light curve obtained in 2006 
has been folded in phase with the frequency 3.49 \cd, with a negative result.
More observations are required to confirm the detected frequency.

The ASAS-3 light curve shows a long-term trend which does not allow us to search for short periods.

%%%%%%%%%%%%%%%%%%%%%%%%%%%%%%%%%%%%%%%%%%%%%%%%%% TERMINADO
\subsubsection{HD 46484} 
This star has been detected as spectroscopic variable by Hubert \citetext{priv.\ comm.}.
No results have been obtained from the Hipparcos data, due to the bad sampling of the data and the few observed points.

\begin{figure}
  \centering
  \includegraphics[width=8cm]{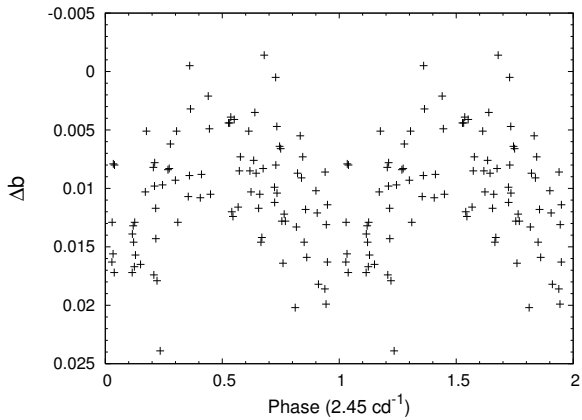}
  \caption{Phase diagram of HD 46484 folded with the frequency $2.45$ \cd\ for the OSN data in the $b$ filter.}
  \label{fig:46484:1}
\end{figure}

Our 4-day light curve obtained in 2004 at the OSN shows variability with an amplitude of the order of 5 mmag. 
A frequency at $2.45$ \cd\ and its 1-day aliases appear in the periodogram of the $vby$ light curves. 
The phase diagram of the $b$ light curve folded with the frequency $2.45$ \cd\ is presented in Fig.~\ref{fig:46484:1}. 
This star is saturated in the ASAS-3 database. 

%%%%%%%%%%%%%%%%%%%%%%%%%%%%%%%%%%%%%%%%%%%%%%%%%%
\subsubsection{HD 48282} 

\begin{figure}
  \centering
  \includegraphics[width=8cm]{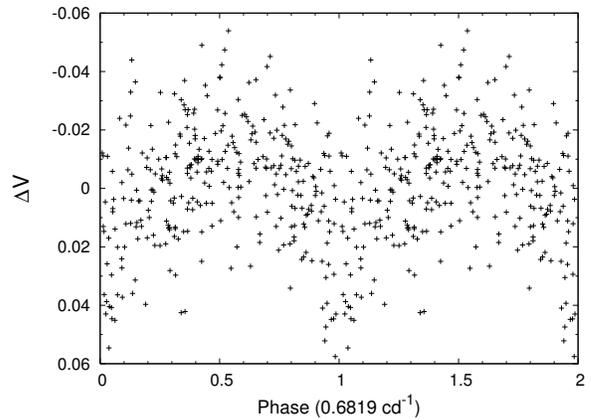}
  \caption{Phase diagram of HD 48282 folded with the frequency $0.6819$ \cd\ for the ASAS-3 data.}
  \label{fig:48282:1}
\end{figure}

A long-term trend is present in the light curve of Hipparcos and ASAS-3, which has been 
removed with a low-order polynomial. No results have been found with the Hipparcos data. 
The frequency analysis of the \mbox{ASAS-3} dataset yields a main variation at $0.6819$ \cd.
The corresponding phase diagram (Fig.~\ref{fig:48282:1}) is very scattered, probably due to the presence of the long-term trend. 

%%%%%%%%%%%%%%%%%%%%%%%%%%%%%%%%%%%%%%%%%%%%%%%%%%%
\subsubsection{HD 49330} 
 
The Hipparcos data show a clear periodicity of $0.283$ d, i.e. at frequency $3.534$ \cd. %poner algo mas!!

This star was observed at the OSN during three observing runs, in 2003, 2004 and 2006.
In 2003, our 8.5-hour light curve presents clear variability. %(upper panel of Fig.\ref{fig.49330:1}) . 
The short-time coverage does not allow us to perform a period search. 
In 2004, our 4-day light curve does not show this variation. Only peaks at $1,2, \ldots$ \cd\ are present in the periodograms
of each filter, with an incoherent phase diagram.
In 2006, a frequency at $1.15$ \cd\ and its 1-day aliases strongly appear in the periodogram. 

\begin{figure}
   \centering
   \includegraphics[width=8cm]{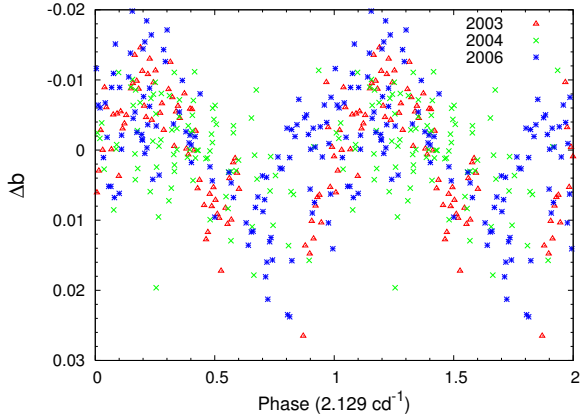} 
   \caption{Phase diagram of the combined dataset obtained at the OSN for star HD 49330 folded with the frequency $2.129$ \cd. 
Triangles in red, crosses in green and asterisks in blue correspond to the 2003, 2004 and 2006 observations respectively.
}
   \label{fig:49330:1}
   \end{figure}

As a final check, we have combined the data of the three seasons, which allows us to detect a frequency at $2.129$ \cd. This frequency is a 1-day alias of the frequency found in the 2006 data.
In Fig.~\ref{fig:49330:1} we show the light curve 
in phase with this frequency for the three seasons with different symbols and colours.
This figure suggests that this frequency has been present in the light curve over the three years, but its amplitude has changed dramatically,
from 13 mmag in 2003 to 4 mmag in 2004 and 8 mmag in 2006 in the $v$ filter. 
Errors in the amplitude are of the order of 1 mmag in this filter. 

The light curve of the ASAS-3 data shows a quasi-cyclical variation of about 1600 days, disturbed by an outburst of shorter 
duration of about 200 days. We noticed that the OSN data obtained in 2003 was observed during this short outburst.

%%%%%%%%%%%%%%%%%%%%%%%%%%%%%%%%%%%%%%%%%%%%%%%%%%%%%
\subsubsection{HD 49585} 

 \begin{figure}
   \centering
   \includegraphics[width=9cm]{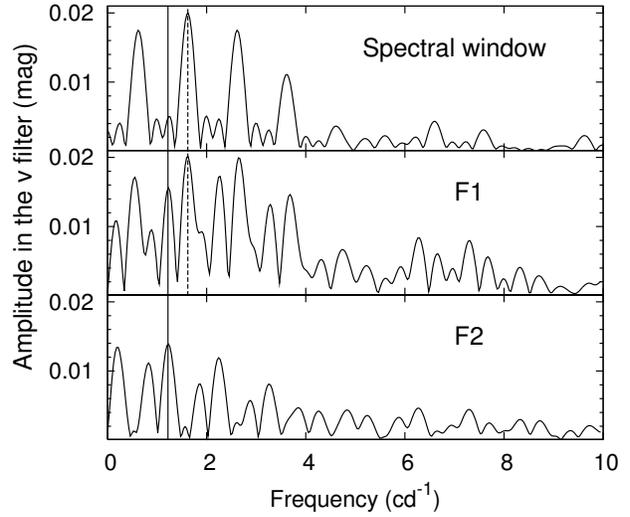}
   \caption{Successive periodograms and spectral window shifted at frequency F1 
of the light curve of HD 49585 in the $v$ filter. 
The dashed and solid vertical arrows stand for the position of frequencies $\mathrm{F}1=1.647$ \cd\  
and $\mathrm{F}2=1.218$ \cd\ respectively. 
}
   \label{fig:49585:1}
   \end{figure}

\begin{figure}
   \centering
   \includegraphics[width=9cm]{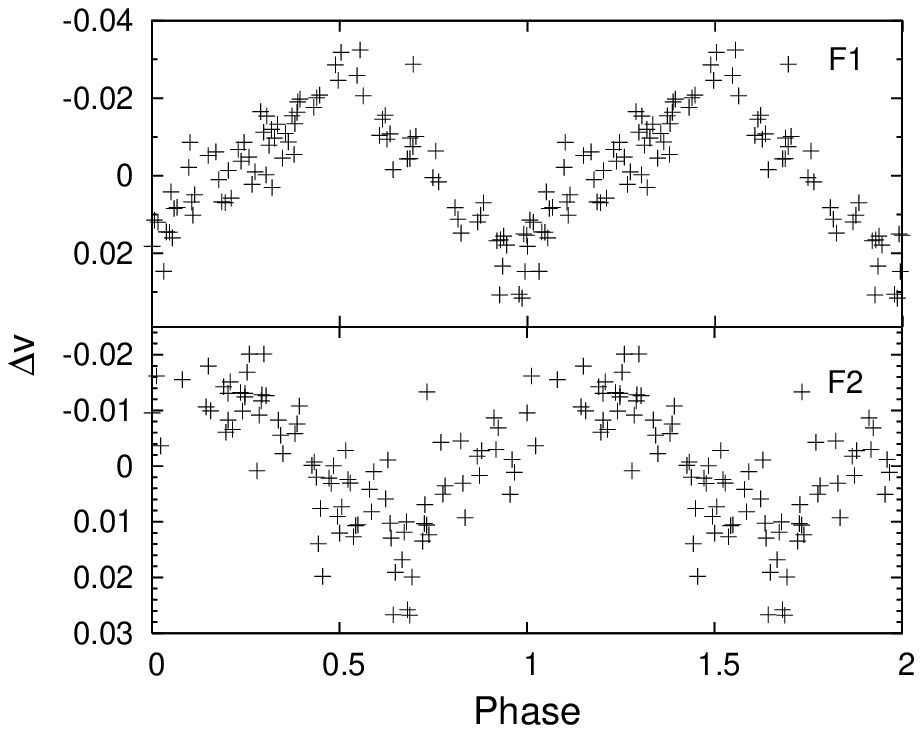}
   \caption{Phase plots of HD 49585 folded with the frequency $\mathrm{F}1=1.647$ \cd, 
after prewhitening for $\mathrm{F}2=1.218$ \cd (\textbf{top}) and 
with the frequency $\mathrm{F}2=1.218$ \cd (\textbf{bottom}), after prewhitening for $\mathrm{F}1=1.647$ \cd.}
   \label{fig:49585:2}
   \end{figure}

There are no Hipparcos data for this faint star.
This star was observed during one season at the OSN in 2004. Only detailed analysis 
for the $v$ filter is provided here, since results with other filters are similar within errors.
The periodogram shows strong peaks at frequency $\mathrm{F}1=1.647$ \cd\ and its 1-day aliases 
(see middle panel of Fig.~\ref{fig:49585:1}). In the top panel of Fig.~\ref{fig:49585:1}, we represent the 
spectral window of the light curve shifted at frequency F1. Another peak is present 
in the data at frequency 
$\mathrm{F}2=1.218$ \cd, which is not seen in the spectral window. %cambiar
Periodogram of the residuals after prewhitening for frequency F1 is depicted in the \textbf{bottom} panel of 
Fig.~\ref{fig:49585:1}. 
The 1-day alias of F2 ($0.21$ \cd), gives a similar fit, although phase diagram for F2 is much better. 
The phase plots for frequencies F1 and F2 are displayed in Fig.~\ref{fig:49585:2}.

%ASAS variable.
A long-term trend is apparent in the light curve of the ASAS-3 dataset, which does not allow us to search for short-term periodicity.

%%%%%%%%%%%%%%%%%%%%%%%%%%%%%%%%%%%%%%%%%%%%%%%%%%%%%
\subsubsection{HD 50087} 
We observed this star in 2006 at the OSN.
Different frequencies are detected in the four filters with very low signal-to-noise ratio, suggesting that these frequencies are artifacts.
The light curve of the ASAS-3 data does not show any significant variation.

%%%%%%%%%%%%%%%%%%%%%%%%%%%%%%%%%%%%%%%%%%%%%%%%%%%%%%
\subsubsection{HD 50209} 

The Hipparcos data analysis yields to a frequency at $1.689$ \cd, considered as uncertain. Another peak 
at frequency $1.47$ \cd\ is also present, although with a lower amplitude.

 \begin{figure}
   \centering
   \includegraphics[width=8cm]{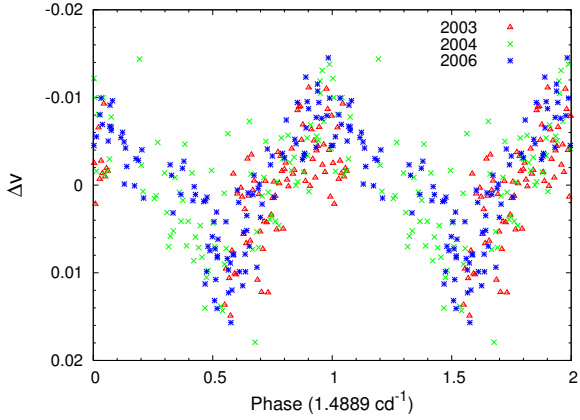}
   \caption{Phase plot of HD 50209 folded with the frequency $1.4889$ \cd\ for the combined data obtained at the OSN.
Triangles in red, crosses in green and asterisks in blue correspond to the 2003, 2004 and 2006 observations respectively.}
   \label{fig:50209:3}
   \end{figure}

 \begin{figure}
   \centering
   \includegraphics[width=8cm]{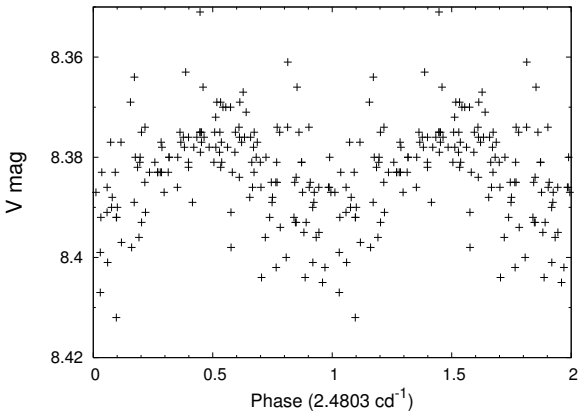}
   \caption{Phase plot of HD 50209 folded with the frequency $2.4803$ \cd\ for the ASAS-3 dataset.}
   \label{fig:50209:4}
   \end{figure}

This star was observed during three observing runs at the OSN. 
The 2003 dataset shows clear variability, but the time span of 9 hours does not allow us to perform a spectral analysis.
In 2004, a frequency at $1.52$ \cd\ is clearly found in all filters, while in 2006 a similar 
frequency at $1.48$ \cd\ is also detected in all filters.
The combined data of the three seasons allows us to refine the frequency to $1.4889$ \cd 
(see Fig.~\ref{fig:50209:3}). Note that we have removed the average magnitude of each year before combining the 
datasets.

The periodogram of the ASAS-3 dataset shows significant peaks at frequency $2.4803$ \cd\ and its 
daily aliases. The phase plot with this frequency is depicted in Fig.~\ref{fig:50209:4}.
This frequency is probably a 1-day alias of the frequency detected at the OSN.
However, the phase diagram of the ASAS-3 dataset with the frequency $1.4889$ \cd\ and the light curve obtained at the 
OSN folded with the frequency $2.4803$ \cd\ are very scattered.
This could be due to the fact that the ASAS-3 dataset contains long- and mid-term trends or to the
presence of multiple periods.
After prewhitening for the frequency $2.4803$ \cd\ in the ASAS-3 dataset, a frequency at $1.4749$ \cd\ appears in the 
periodogram, but the phase diagram with this frequency is very scattered.
More observational data is required to confirm the multiperiodicity of this star.

%%%%%%%%%%%%%%%%%%%%%%%%%%%%%%%%%%%%%%%%%%%%%%%%%%%%%%
\subsubsection{HD 50696} 

\begin{figure}
   \centering
   \includegraphics[width=8cm]{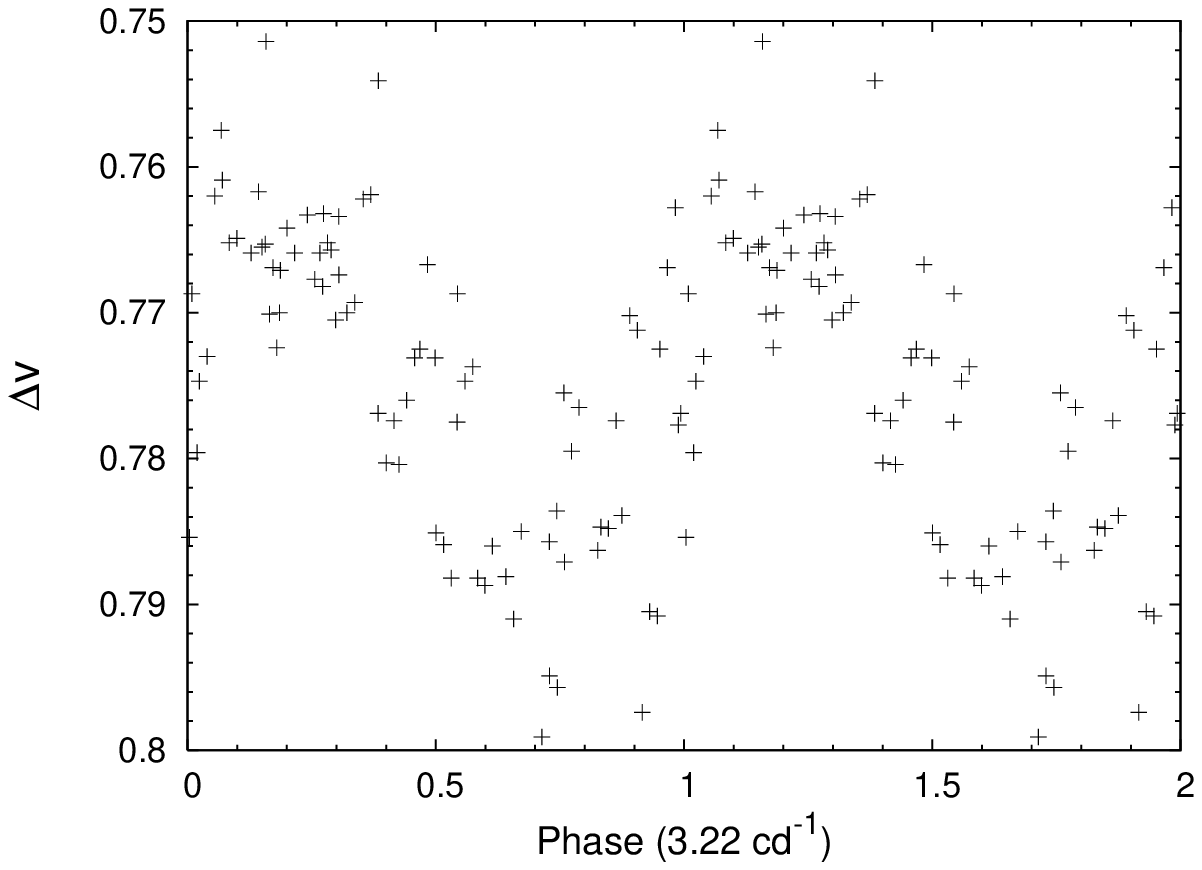}
   \caption{Phase plot of HD 50696 folded with the frequency $3.22$ \cd\ for the OSN data obtained in 2004.}
   \label{fig:50696:1}
   \end{figure}

There are no Hipparcos data for this faint star. 
We observed this source during two seasons at the OSN.
The 2003 dataset shows clear variability, but the time span does not allow us to perform any period analysis. 
A clear frequency at $3.22$ \cd\ is detected in the 2004 photometry through the four filters.
We present the phase diagram, only for the $v$ filter data for clarity, in Fig.~\ref{fig:50696:1}. 
Note that the amplitude of the variation is of the order of 12 mmag.
The phase diagram of the data obtained in 2003 is compatible with this frequency.
 
A long-term variation is found in the ASAS-3 dataset, which stops us 
from searching for short-term variability.

%%%%%%%%%%%%%%%%%%%%%%%%%%%%%%%%%%%%%%%%%%%%%%%%%%%%%%
\subsubsection{HD 50891} 
\begin{figure}
   \centering
   \includegraphics[width=8cm]{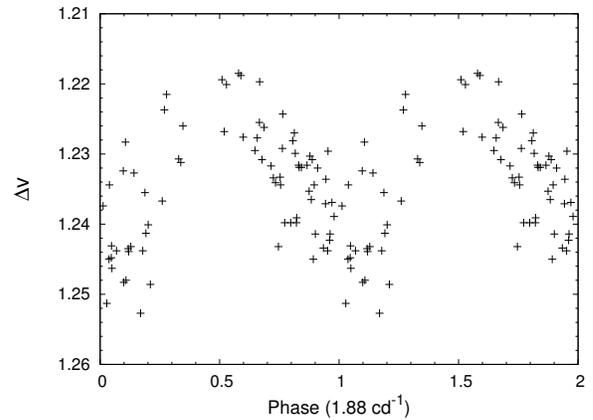}
   \caption{Phase plot of HD 50891 folded with the frequency $1.88$ \cd\ for the OSN data obtained in 2004.}
   \label{fig:50891:1}
   \end{figure}

There are no Hipparcos data for this star.
This star has been observed only in 2004 at the OSN.
The periodogram shows significant peaks at frequency $1.88$ \cd\ and its 1-day aliases for all the filters. 
We present the phase curve with this frequency in Fig.~\ref{fig:50891:1}.

A long-term trend is apparent in the light curve of the ASAS-3 dataset, which does not allow us to search for short-term periodicity.

%%%%%%%%%%%%%%%%%%%%%%%%%%%%%%%%%%%%%%%%%%%%%%%%%%%%%% 
\subsubsection{HD 51193} 

\begin{figure}
  \centering
  \includegraphics[width=8cm]{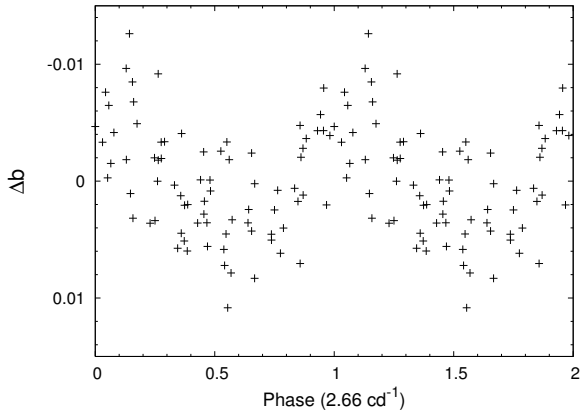}
  \caption{Phase plot of HD 51193 folded with the frequency $2.66$ \cd\ in the $b$ filter for the OSN data obtained in 2004.}
  \label{fig:51193:1}
\end{figure}

\begin{figure}
  \centering
  \includegraphics[width=8cm]{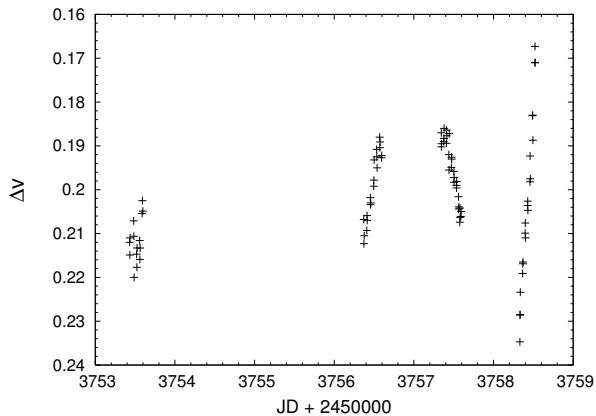}
  \caption{Light curve of HD 51193 for the OSN data obtained in 2006. Note the variation of the amplitude from night to night.}
  \label{fig:51193:2}
\end{figure}

The Hipparcos light curve shows clear periodicity at frequency $1.639$ \cd. 

This star has been observed at the OSN in 2004 and 2006.
A long-term trend is present in the light curve of 2004, which we have removed with a low-order polynomial %mirar 
function. We have found a frequency at $2.66$ \cd\ in the $vby$ filters, which is similar to a 1-day alias of the Hipparcos frequency. 
Both frequencies have a similar fit for the OSN data, although the phase diagram with the frequency $2.66$ \cd\ is much better
(see Fig.~\ref{fig:51193:1}).
In 2006, the amplitude seems to 
vary from night to night, as shown in Fig.~\ref{fig:51193:2}.
Unfortunately, we could not find any significant frequency with a coherent phase diagram. 

The light curve obtained with the ASAS-3 project shows a long-term trend. 
After prewhitening for this trend with a low-order polynomial, a peak at frequency 
$1.6060$ \cd\ and its daily aliases appear in the periodogram. 
The frequency fulfils the SNR criterion, but the phase diagram is very scattered.
Note that this frequency is similar to the frequency detected in the Hipparcos data and the 1-day alias of the 
frequency detected at the OSN.

%%%%%%%%%%%%%%%%%%%%%%%%%%%%%%%%%%%%%%%%%%%%%%%%%%%%%%
\subsubsection{HD 51404}

\begin{figure}
  \centering
  \includegraphics[width=9cm]{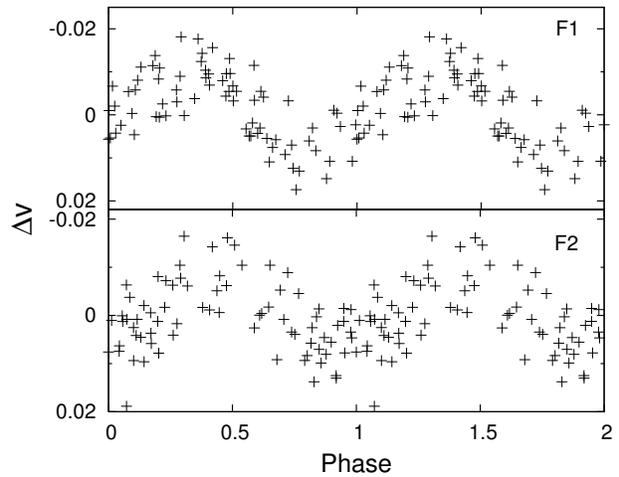}
  \caption{Phase plots of HD 51404 folded with the frequency $\mathrm{F}1=2.68$ \cd\ after prewhitening for $\mathrm{F}2=5.99$ \cd 
 (\textbf{top}) and 
with the frequency $\mathrm{F}2=5.99$ \cd\ after prewhitening for $\mathrm{F}1=2.68$ \cd\ (\textbf{bottom}) in the $v$ filter of the 2004 dataset.}
  \label{fig:51404:1}
\end{figure}

There are no Hipparcos data for this star.
This star was only observed in 2004 at the OSN. 
It is clearly variable with an amplitude of 10 mmag. 
Strong peaks appear in the periodogram of the $vby$ filters at frequency $\mathrm{F}1=2.68$ \cd\ and its daily aliases. 
The light curve folded with this frequency is depicted in Fig.~\ref{fig:51404:1} (upper panel).
After prewhitening for this frequency, another significant frequency at $\mathrm{F}2=5.99$ \cd\ is detected, also 
in the $vby$ filters.
The phase plot with this frequency is displayed in the lower panel of Fig.~\ref{fig:51404:1}, only for the $v$ filter data for clarity.
%tabla con amplitudes?

The ASAS-3 light curve shows short- and long-term variability with a standard deviation of the observations of 24 mmag.
However, no significant frequencies are found.

%%%%%%%%%%%%%%%%%%%%%%%%%%%%%%%%%%%%%%%%%%%%%%%%%%%%%%
\subsubsection{HD 51452} 

\begin{figure}
  \centering
  \includegraphics[width=8cm]{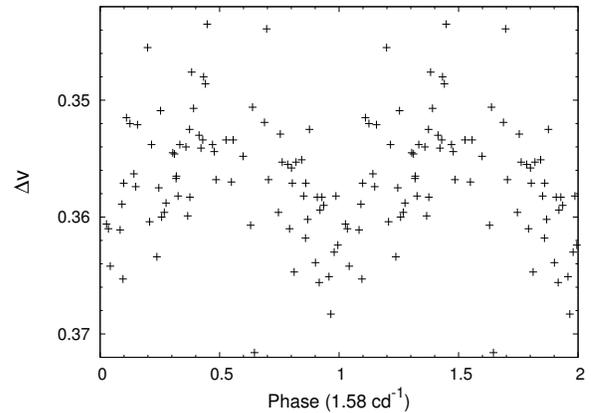}
  \caption{Phase plot of HD 51452 folded with the frequency $1.58$ \cd\ in the $v$ filter for the OSN data.} 
  \label{fig:51452:1}
\end{figure}

There are no Hipparcos data for this star.
We observed this star at the OSN in 2004.
The light curve shows variability with the frequency $1.58$ \cd, but only in the $vy$ filters,  
and thus, this frequency is considered as uncertain.
The phase diagram with this frequency is
displayed in Fig.~\ref{fig:51452:1} for filter $v$.
The ASAS-3 data does not show any indication of variability.

%%%%%%%%%%%%%%%%%%%%%%%%%%%%%%%%%%%%%%%%%%%%%%%%%%%%%%%%%%%%%%%
%%%%%%%%%%%%%%%%%%%%%%%%%%%%%%%%%%%%%%%%%%%%%%%%%%%%%%%%%%%%%%%
%%%%%%%%%%%%%%%%%%%%%%%%%%%%%%%%%%%%%%%%%%%%%%%%%%%%%%%%%%%%%%%
%%%%%%%%%%%%%%%%%%%%%%%%%%%%%%%%%%%%%%%%%%%%%%%%%%%%%%%%%%%%%%%
%%%%%%%%%%%%%%%%%%%%%%%%%%%%%%%%%%%%%%%%%%%%%%%%%%%%%%%%%%%%%%%

\subsection{Stars in the Galactic Centre Direction}

%%%%%%%%%%%%%%%%%%%%%%%%%%%%%%%%%%%%%%%%%%%%%%%%%%%%%%
\subsubsection{HD 170\,714}
The few datapoints collected by the Hipparcos mission for this star do not allow us to search for 
short-term variability.
 
\begin{figure}
  \centering
  \includegraphics[width=8cm]{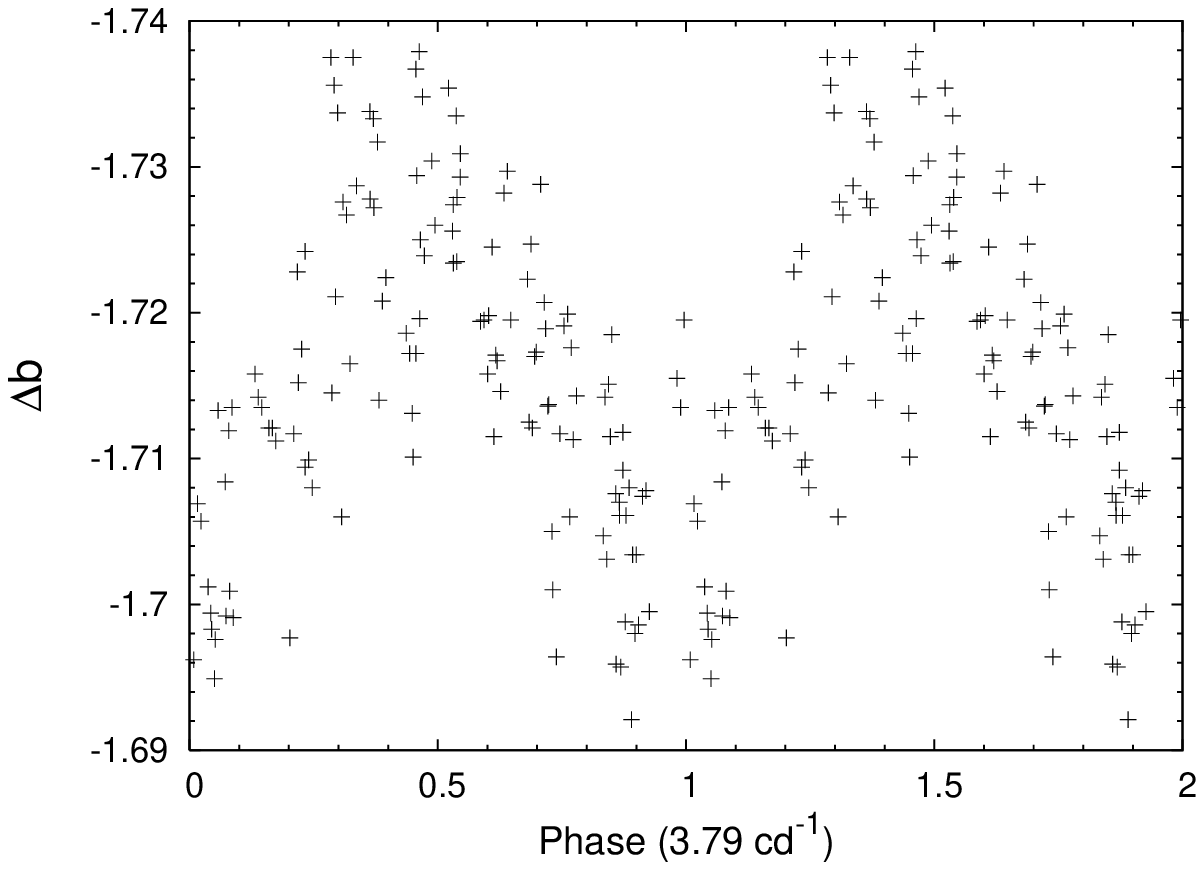}
  \caption{Phase plot of HD 170\,714 folded with the frequency $3.79$ \cd\ in the $b$ filter for the OSN data.} 
  \label{fig:170714:1}
\end{figure}

This star was observed at the OSN in 2005. Strong peaks at frequency $3.79$ \cd\ and its 1-day aliases 
appear in the periodogram of all the filters. The amplitude corresponding to this frequency is 10 mmag 
for the $vby$ filters ($\pm 1$ mmag) and 13 for the $u$ filter ($\pm 2$ mmag). 
In Fig.~\ref{fig:170714:1} we represent the light curve folded in phase with the detected frequency 
for the $b$ filter.

This star is saturated in the ASAS-3 dataset.

%%%%%%%%%%%%%%%%%%%%%%%%%%%%%%%%%%%%%%%%%%%%%%%%%%%%%%
\subsubsection{HD 171\,219} 
This star is classified as constant in the Hipparcos catalogue.
We have re-analysed the light curve and no short-term variability has been found. 
We observed this star at the OSN during two seasons, in 2002 and 2003. 
The periodograms of both datasets do not show any significant peak greater than 2 mmag.
We conclude that the star is not variable at our detection level. This star is saturated in the ASAS-3 database.

%%%%%%%%%%%%%%%%%%%%%%%%%%%%%%%%%%%%%%%%%%%%%%%%%%%%%%
\subsubsection{HD 173\,292} 
The Hipparcos data show short-term variability with a standard deviation of 30 mmag. 
However, no significant frequencies have been detected in the Fourier analysis.

\begin{figure}
  \centering
  \includegraphics[width=8cm]{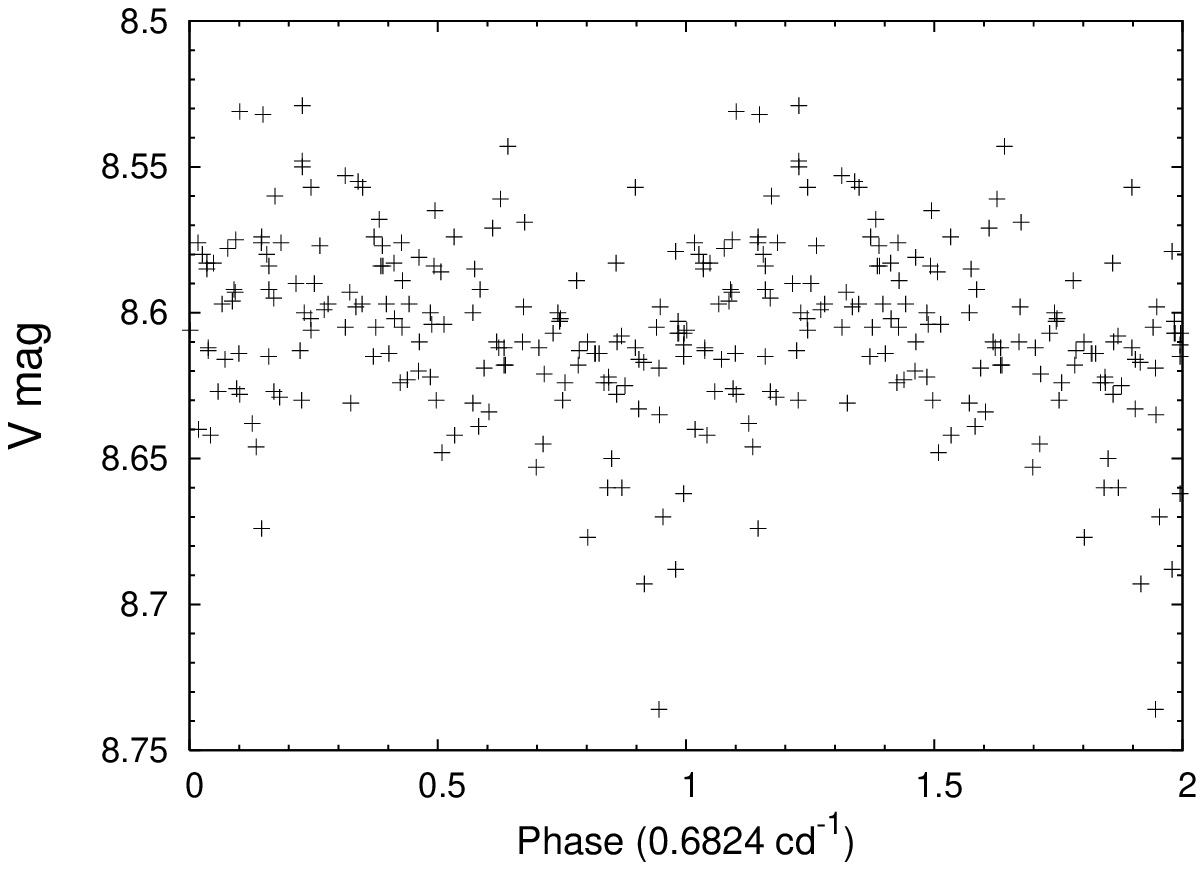}
  \caption{Phase plot of HD 173\,292 folded with the frequency $0.6824$ \cd\ for the ASAS-3 dataset.} 
  \label{fig:173292:1}
\end{figure}

The spectral analysis of the ASAS-3 dataset shows peaks at frequency  
$0.6824$ \cd\ and its daily aliases. A phase diagram with this frequency 
is depicted in Fig.~\ref{fig:173292:1}. Note the high amplitude of the variation ($\sim$16 mmag).

This star was not observed at the OSN.

\begin{figure}
  \centering
  \includegraphics[width=8cm]{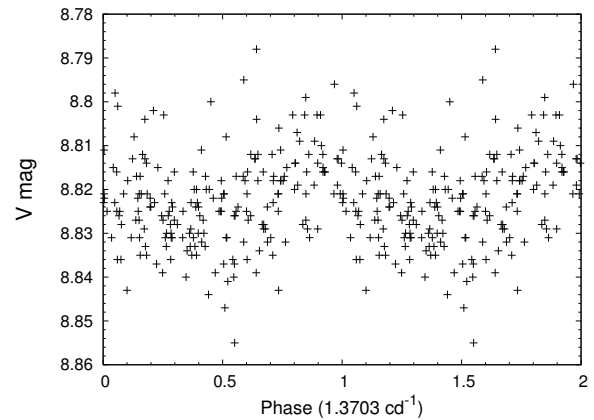}
  \caption{Phase plot of HD 173\,530 with the frequency $1.3703$ \cd\ for the ASAS-3 dataset.} 
  \label{fig:173530:1}
\end{figure}

%%%%%%%%%%%%%%%%%%%%%%%%%%%%%%%%%%%%%%%%%%%%%%%%%%%%%%
\subsubsection{HD 173\,530} 

There are no Hipparcos data for this faint star.
A significant peak at 
frequency $1.3703$ \cd\ appears in the periodogram of the ASAS-3 light curve with an amplitude of 6 mmag.
A phase diagram for this frequency is plotted in Fig.~\ref{fig:173530:1}.

This star has not been observed at the OSN.

%%%%%%%%%%%%%%%%%%%%%%%%%%%%%%%%%%%%%%%%%%%%%%%%%%%%%%
\subsubsection{HD 173\,637} 

The Hipparcos data show a long-term variation, which does not allow us to search for short-term periodicity.

\begin{figure}
  \centering
  \includegraphics[width=8cm]{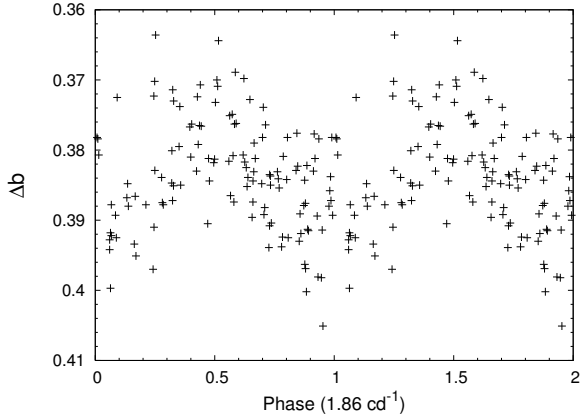}
  \caption{Phase plot of HD 173\,637 folded with the frequency $1.86$ \cd\ in the $b$ filter for the OSN data.} 
  \label{fig:173637:1}
\end{figure}

This star was observed at the OSN during two seasons, in 2003 and 2005. However, 
only 19 datapoints were obtained in 2005 and therefore no result can be achieved with this dataset.
Short-term variability is found in the 2003 dataset.
We detect significant peaks in the frequency domain at $1.86$ \cd\ and its 1-day aliases. 
The amplitude of this frequency is of the order of 5 mmag in the $vby$ filters and the SNR is 4.8.
The phase diagram with this frequency in the $b$ filter is displayed in Fig.~\ref{fig:173637:1}.

\begin{figure}
  \centering
  \includegraphics[width=8cm]{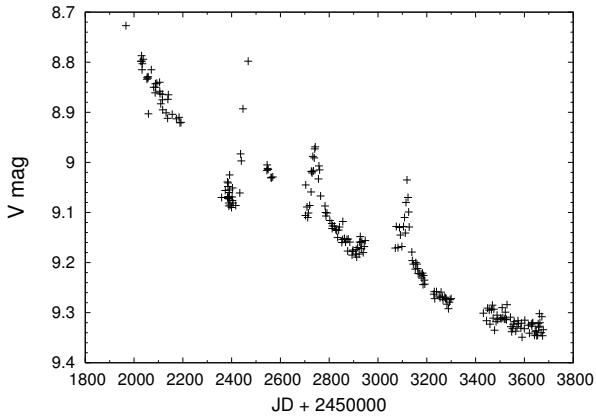}
  \caption{Light curve of HD 173\,637 for the ASAS-3 dataset.}
  \label{fig:173637:2}
\end{figure}

A long-term trend is 
present in the ASAS-3 light curve (Fig.~\ref{fig:173637:2}), superimposed on which we see
several outburst (at least 3) with an amplitude of 0.1 mag and a time difference of around
360 to 380 days. We have not searched for short-term periodicity because of the complexity of the light curve.

%%%%%%%%%%%%%%%%%%%%%%%%%%%%%%%%%%%%%%%%%%%%%%%%%%%%%%
\subsubsection{HD 173\,817} 
The Hipparcos light curve does not show any indication of variability.

This star was observed at the OSN in 2003.
The light curve shows short-term variability with the frequency $3.51$ \cd\ in the $vby$ filters, which fulfils the SNR criterion.
However, the amplitude is of the order of 3 mmag and thus, the phase curve is very 
scattered. 

%ASAS
A long-term trend is present in the ASAS-3 light curve. We do not find short-term variability after 
removing the long-term variation. 
The frequency found at the OSN is not detected in the ASAS-3 dataset, probably due to the fact that
the ASAS-3 photometry have lower precision. 
More observational data are required to confirm the periodicity of this star.

%%%%%%%%%%%%%%%%%%%%%%%%%%%%%%%%%%%%%%%%%%%%%%%%%%%%%%
\subsubsection{HD 174\,513} 
\begin{figure}
\centering
  \includegraphics[width=9cm]{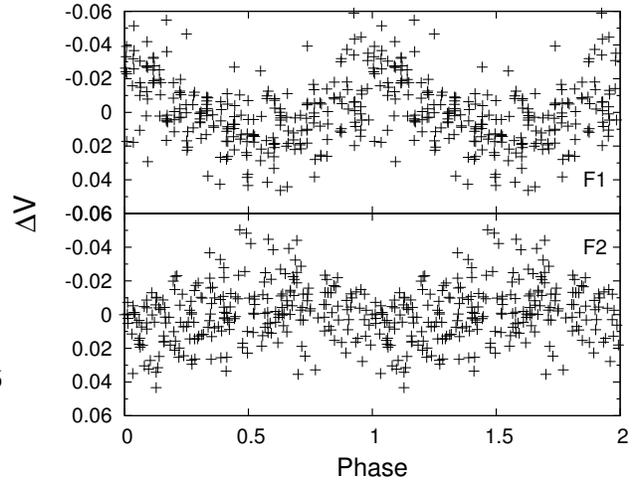}
  \caption{Phase plot of HD 174\,513 folded with the frequency $\mathrm{F}1=0.0271$ \cd, after prewhitening for 
 $\mathrm{F}2=5.293$ \cd (\textbf{top}) and with the frequency $\mathrm{F}2=5.293$ \cd, after prewhitening for $\mathrm{F}1=0.0271$ \cd (\textbf{bottom})
 for the ASAS-3 dataset.}
  \label{fig:174513:1}
\end{figure}

Long- and short-term variations are found in the Hipparcos light curve. 
However, the frequency analysis 
does not yield to any frequency with a coherent phase diagram.

The analysis of the dataset obtained at the OSN in 2002 
confirms its variability with frequencies at $3.34$ \cd\ and its 1-day aliases. 
However, this frequency is very uncertain due to the low amplitude (4 mmag) and the few observed datapoints.

In the ASAS-3 dataset, a long-term trend is present, which have been removed with a low-order polynomial. 
The periodogram of the residual 
shows a strong peak at the low-frequency $\mathrm{F}1=0.0271$ \cd (i.e. 36.90 d) and its daily aliases. A phase plot 
for this frequency is displayed in the upper panel of Fig.~\ref{fig:174513:1}. 
This frequency is too \textbf{low} to be produced by pulsations and it is probably caused by a 
binary component.
Prewhitening for this 
frequency, we found an additional significant frequency at $\mathrm{F}2=5.293$ \cd with an amplitude of 8 mmag, which is similar to a 1-day alias 
of frequency found at the OSN (3.34 \cd). A phase diagram of this frequency is plotted in the lower panel of  Fig.~\ref{fig:174513:1}.
More photometric data are required to confirm the frequency F2.%poner S/N

%%%%%%%%%%%%%%%%%%%%%%%%%%%%%%%%%%%%%%%%%%%%%%%%%%%%%%
\subsubsection{HD 174705} 
\begin{figure}
  \centering
  \includegraphics[width=8cm]{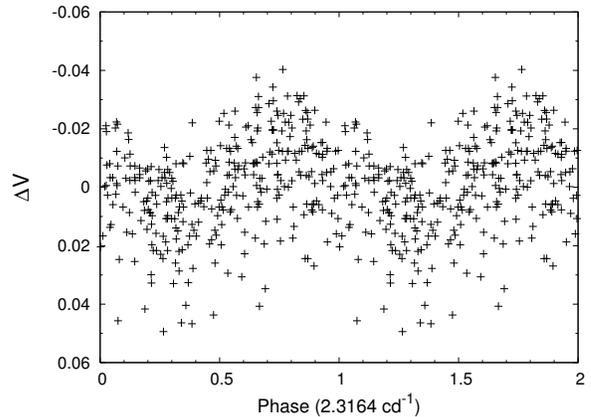}
  \caption{Phase plot of HD 174\,705 folded with the frequency $2.3164$ \cd\ for the ASAS-3 dataset.} 
  \label{fig:174705:1}
\end{figure}
 
There is neither Hipparcos nor OSN data for this star.
The ASAS-3 light curve contains individual subsets with different mean magnitudes due to the fact that 
the star were observed in several different fields, as pointed out by \citet{pojmanski02}. 
Therefore, the mean magnitude has first been subtracted from each subset and then the subsets have been combined. 
A significant peak appears at frequency $2.3164$ \cd in the periodogram of the combined dataset.
The light curve folded in phase with this frequency is displayed in Fig.~\ref{fig:174705:1}.

%%%%%%%%%%%%%%%%%%%%%%%%%%%%%%%%%%%%%%%%%%%%%%%%%%%%%%
\subsubsection{HD 181231} 
The Hipparcos data do not show any indication of variability.
We observed this star at the OSN in 2003.
Unfortunately, the comparison star HD 182\,786 showed variability during the observing run, 
and thus no reliable result can be obtained with this comparison star. Using the check star HD 182\,198, a 
significant peak appears in the periodogram of all filters at frequency $0.67$ \cd, but the phase coverage is not complete. 
In addition, we have analysed photometric data obtained by Ennio Poretti at San Pedro Martir (SPM).
HD 181414 was used in the differential photometry as the comparison star. Frequency 
$0.67$ \cd\ is also detected in the $vby$ filters. The amplitude associated to this frequency is 4 and 3 mmag 
at the OSN and SPM respectively. 

We also performed a detailed analysis of the ASAS-3 light curve, which shows 
short-term variability with the frequency $3.4304$ \cd. However, this frequency is 
considered as uncertain due to the low value of the associated amplitude (5 mmag).
The ASAS-3 dataset has also been plotted in phase with the frequency $0.67$ \cd, resulting in a 
scattered diagram.
The periodogram of the combined data (OSN, SPM, ASAS-3) yields a frequency of $0.6425$ \cd. However, the phase 
diagram has a large scatter and therefore we cannot confirm this frequency.

%%%%%%%%%%%%%%%%%%%%%%%%%%%%%%%%%%%%%%%%%%%%%%%%%%%%%%
\subsubsection{HD 181\,367} 

There are no Hipparcos data for this star.
No short-term variability is found in the light curve obtained at the OSN in 2003 with an amplitude exceeding 4 mmag.
The same conclusions have been reached by Ennio Poretti \citetext{priv.\ comm.} in the 
analysis of the SPM data for this star.

The ASAS-3 light curve shows a long-term trend. After removing it with a low-order polynomial, 
we do not find any indication of variability.

%%%%%%%%%%%%%%%%%%%%%%%%%%%%%%%%%%%%%%%%%%%%%%%%%%%%%%
\subsubsection{HD 183\,656} 

HD 183\,656 is a spectroscopic binary shell star (SB1) with an orbital period of 214.75 d \citep{koubsky89}.
This star was found variable by \citet{lynds60}, who proposed a period of 0.8518 d (i.e. $1.1740$ \cd). 
Short-term variability with a period of 0.652 d (i.e. $1.534$ \cd) was obtained from the Hipparcos photometry \citep{hubert98}.

\begin{figure}
  \centering
  \includegraphics[width=8cm]{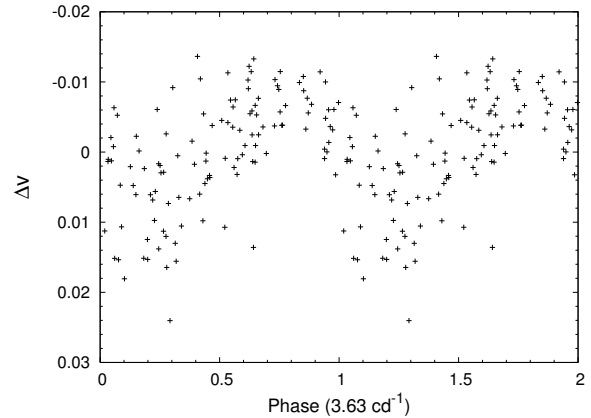}
  \caption{Residuals of HD 183\,656 in phase with the frequency  $3.63$ \cd, after removing the long-term trend.}  
  \label{fig:183656:1}
\end{figure}

We observed this star at the OSN in 2002 and 2005.
In the periodogram of the 2002 light curve, the most powerful peak appears at frequency
$2.19$ \cd, although the phase diagram is very scattered.
Note that the 1-day alias of this frequency is Lynds' period.
However, the Hipparcos data appear very noisy when folded with any of the two frequencies.
The 10-day light curve obtained in 2005 shows a long-term trend, which has been removed with a low-order
polynomial function. 
After pre\-whiten\-ing for this trend, a significant peak appears at frequency $3.63$ \cd\ in the periodogram of the 
light curve in the $vby$ filters with an amplitude of 7 mmag (Fig.~\ref{fig:183656:1}).
The $2.19$ \cd\ frequency is not detected in the 2005 data.
This star is saturated in the ASAS-3 database.

%%%%%%%%%%%%%%%%%%%%%%%%%%%%%%%%%%%%%%%%%%%%%%%%%%%%

\subsubsection{HD 184\,279}

\citet{percy02} performed an autocorrelation analysis of the Hipparcos data for this star and
found a period of $0.6$ days (i.e. $1.667$ \cd), which they presented as uncertain.
A re-analysis of the Hipparcos data with the methods explained above yields a period of 0.156 d (i.e. $6.410$ \cd), 
although the phase diagram is very scattered.

We observed this star at the OSN in 2002 and 2005.
The periodogram of the 2002 dataset shows a peak at frequency $1.38$ \cd, but we cannot confirm this frequency
due to the few points of the sample.
The 3-night light curve obtained in 2005 confirms the presence of short-term variability for this star.
However, frequencies found in the spectral analysis have very scattered phase diagrams and thus are also
very uncertain. 
The long-term trend which is present in the ASAS-3 dataset does not allow us to 
search for short-period variability. 
We conclude that this star is variable, but more observations are needed to confirm its periodicity.

\end{document}